\documentclass[pra,twocolumn,aps,showpacs,amsmath,amssymb]{revtex4}

\usepackage[english]{babel}

\usepackage[latin1]{inputenc}

\usepackage{graphicx}

\usepackage{latexsym}

\usepackage{graphics}

\usepackage{epsfig}

\usepackage{amssymb}

\def\refer#1{(\ref{#1})}

\def\im{ {\rm{i}} }

\def\r{ {\bf{r}} }

\def\R{ {\bf{R}} }

\def\dx{ {\rm{d}x} }

\def\dy{ {\rm{d}y} }

\def\kago{kagom\'{e}}

\def\be{\begin{equation}}          \def\ee{\end{equation}}

\def\bea{\jot2ex\begin{eqnarray}}         \def\eea{\end{eqnarray}}

\def\befi{\begin{figure}[5mm]} \def\efi{\end{figure}}

\def\re#1{\ref{#1}}

\def\t{\tau}

\def\hta{\hat{\tau}}

\def\oa{\omega^A} \def\ob{\omega^B}  \def\of{\omega^{\rm ferro}}

\def\d{ {\rm{d}} }

\def\k{ {\bf{k}} }

\def\ex#1{ {\rm{e}}^{#1} }

\def\W{\mathcal{W}}

\begin{document}

\title{Quantum gases in trimerized kagom\'e lattices}

\author{B. Damski$^{1,2}$, 
H. Fehrmann$^{1}$, 
H.-U. Everts$^{1}$, 
M. Baranov$^{3,4}$, L. Santos$^{5}$, 
and M. Lewenstein$^{1,6}$\footnote{also at Instituci\`o Catalana de Recerca i Estudis Avan\c cats.}}  

\affiliation{
(1) Institut f\"ur Theoretische Physik, Universit\"at Hannover, D-30167 Hannover, Germany\\ 
(2) Theory Division, Los Alamos National Laboratory, MS-B213, Los Alamos,
87545 NM, USA\\ 
(3) Van der Waals-Zeeman Instituut, 
Universiteit van Amsterdam, 1018 XE Amsterdam, The Netherlands\\ 
(4) Kurchatov Institute, Kurchatov sq. 1, 123182 Moscow, Russia\\
(5) Institut f\"ur Theoretische Physik III, Univesit\"at Stutgart, 
D-70550 Stuttgart, Germany\\ 
(6) ICFO-Institut de Ci\`encies Fot\`oniques, Jordi Girona 29, Edifici Nexus II, E-08034 Barcelona, Spain}

\begin{abstract}

We study low temperature properties of atomic gases in
trimerized optical kagom\'{e} lattices. The laser arrangements 
that can be used to create these lattices are briefly described. We 
also present explicit results for the coupling constants of the generalized 
Hubbard models that can be realized in such lattices. 
In the case of  a single component Bose gas the
existence of a Mott insulator phase with fractional numbers of particles per trimer 
is verified in a mean field approach.   
The main emphasis of the paper is on an atomic 
spinless interacting Fermi gas in the trimerized kagom\'{e} lattice 
with two fermions per site. This system is shown to be described 
by  a quantum spin 1/2 model on the triangular lattice with couplings 
that depend on the bond directions. We investigate this model by means 
of exact diagonalization. Our key finding is that the system 
exhibits non-standard properties of a {\em quantum spin-liquid 
crystal}: it combines planar antiferromagnetic order in the ground 
state with an exceptionally large number of low energy excitations. 
The possibilities of experimental verification of our theoretical 
results are critically discussed.  

\end{abstract}

\maketitle

\section{Introduction}
\label{sec:Introduction}

The experimental realization of the Bose-Einstein condensation (BEC)
\cite{bec} linked the physics of cold atoms with that 
of weakly interacting many-body systems, traditionally studied by condensed
matter physics. More recently, the seminal theory paper 
by Jaksch {\it et al.} \cite{Jaksch}, followed by equally seminal experiments by Greiner {\it et al.} 
\cite{Greiner} on the Mott insulator (MI) to superfluid (SF) transition have
paved the way towards the analysis of strongly correlated 
systems within the physics of cold atoms. 
In this sense, the physics of cold atoms is nowadays merging with condensed matter physics, 
solid state physics, and quantum information at the same common frontiers and open challenging problems, 
such as for instance BEC-BCS crossover (see for instance \cite{becbcs}), fractional quantum Hall effect 
(c.f. \cite{fqhe}), physics of 1D systems \cite{tonks}, etc. Quantum information has given new impulses towards 
the understanding of quantum phase 
transitions \cite{qi-qpt},  and to understand better the known (and develop new) numerical methods of 
treating many-body systems \cite{qi-dmrg}.

\subsection{Atomic lattice gases}

One of the most fascinating playgrounds of the cold atom physics is provided
by ultra cold lattice gases, i.e. cold atoms trapped in optical lattices
produced by standing laser waves where, in the case of red (blue) detuned
lasers light, the potential minima coincide with the
intensity maxima (minima) \cite{Grimm}. This technique  
has been of enormous interest
during the past years. The setup can be chosen to be one-, two- or three-dimensional, where the
lattices form ranges from simple periodic (such as a square in 2D, or cubic lattice in  3D, respectively) 
 to more exotic lattices, such as hexagonal \cite{Duan}, or kagom\'e \cite{kagome}
lattices, created with the use of superlattice techniques \cite{Superlattice}.
In experiments, optical lattices  offer an unprecedentedly   wide range of tunable parameters, 
 which can be changed during the evolution {\it in situ} and ``{\it in vivo}'', i.e. in real time. 
These possibilities, on one hand,  link strongly  the physics of ultracold atoms in optical
 lattices to various areas of condensed matter physics, and on the other, they open completely 
new ways to study quantum many-body systems, to perform  in various ways quantum information processing 
(cf. \cite{jaksch2,blochinf}), and even to realize special 
purpose quantum computers, so-called quantum simulators \cite{scala}.

The physics of ultracold atomic gases in optical lattices is in general described by various 
versions of the Hubbard model, which is probably the most important and structurally simple 
model of condensed matter physics, capable nevertheless to describe an enormous variety
of physical phenomena and effects \cite{auer94,Frahm}. 
Atomic ultracold gases may serve as a ``Hubbard model tool-kit''
\cite{jakschzoller}, and several models have been discussed in more detail in this context:
the most simple Bose Hubbard model \cite{Jaksch} (for the seminal condensed matter 
treatment see \cite{fisher}), the Fermi-Fermi model (which should eventually allow for quantum 
simulations of high $T_c$ superconductivity \cite{hof}) the Fermi-Bose model (which leads 
to creation of composite fermions via fermion-boson, or fermion-bosonic hole pairing, c.f. 
\cite{Lewenstein,fehrmann}), or Bose-Bose, or more generally multicomponent systems. 
Quenched disorder may be introduced in a controlled way to such systems \cite{boseglass,sg}, which 
opens the possibility of studying the physics of disordered systems in this framework. 

In a certain limit Hubbard models reduce to spin models and this possibility has been 
also intensively investigated recently for both atomic gases \cite{Duan,spins,mlzo},
and ion chains \cite{ions}. Spin models enjoy particular interest because of
their simplicity and thus possible 
applicability in quantum information processing (c.f. \cite{mlzo,briegel}). In this paper we will 
discuss yet another possibility, i.e. the possibility of studying frustrated 
quantum antiferromagnets (AFM).

\subsection{Quantum antiferromagnets}

Quantum antiferromagnets, and in particular frustrated AFM's are in the center
of interest of modern condensed matter physics 
(for a  recent review, see \cite{lhuillier}). One of the reasons for this interest is that 
frustrated AFM´s are believed 
to explain certain aspects of the high $T_c$ superconductivity \cite{auer94}. 
In this context  frustrated spin-$1/2$ models have attracted particular attention.
At the same time almost all of these models are notoriously  difficult to handle 
analytically and numerically. The only exceptions are those models that
exhibit long range N\'eel type order, since 
there is a powerful method by which long range order can be identified numerically 
(see e.g. \cite{LhuiSinFou01}), and if it exists, the semiclassical (spin-wave) approximation yields 
satisfactory results. In 2D only very few exactly solvable  spin-$1/2$ models are known 
\cite{lhuillier}. 
In 1D exact results can be obtained by the Bethe-Ansatz technique in a number of cases 
\cite{suth}. Moreover non-perturbative bosonisation techniques and powerful 
numerical methods such as the Density Matrix Renormalization Group (DMRG) method can be applied.
However, DMRG techniques become very difficult to handle in the 
case of disordered systems \cite{schollweck}.  In 2D, in the absence of long-range order,
apart from renormalization group approaches
 numerical methods offer the only possibility to investigate  frustrated spin systems.
However, 
Quantum Monte Carlo (QMC) simulations of Heisenberg AFMs on frustrated
lattices, such as the triangular and the kagom\'e lattice,  suffer from  
the ``negative sign'' problem. 
For instance, attempts to obtain useful results for the Heisenberg AF on a triangular
lattice (TAF) by QMC have been futile as a consequence of this problem.
Because of the experience with this and 
other frustrated models, we expect that the ``negative sign'' problem also invalidates QMC 
for the system to be studied in this paper. In contrast with the failure of QMC, 
exact diagonalization of the Hamiltonian of the TAF for rather small cells of the lattice 
has produced the main result for this model:
its ground state, contrary to earlier conjectures, was shown to be long range ordered 
(see \cite{bern}).

According to C. Lhuillier and her collaborators quantum Heisenberg AFM's at very low temperatures  
exhibit 4 distinct kinds of quantum phases:
\begin{itemize}
\item{ Semiclassical ordered N\'eel phases, characterized by long range order 
in spin-spin correlation function, 
breaking of the $SU(2)$ symmetry, and gapless spectrum with $\Delta S_z=1$
 magnon excitations. The standard example of such order is 
provided by the Heisenberg AFM on a square lattice in 2D. 
The theoretical description of such systems using the spin wave theory 
(cf. \cite{auer94}) is quite accurate. }
\item{Valence Bond Crystals (VBC) (or Solids), characterized by 
long range order in dimer coverings, with prominent examples being the AKLT
model 
in 1D \cite{aklt}, or the Heisenberg model on the 2D checker-board lattice
\cite{lhuillier,StaFuBa05} (corresponding to a 2D slice of a pyrochlore
lattice). VBC's exhibit no $SU(2)$ symmetry breaking,
short range spin-spin correlations, long range dimer-dimer order and/or order or long range order 
 of larger $S=0$ plaquettes, 
and gapped excitations in all $S$ sectors.
}
\item{Resonating Valence Bond spin liquids (Type I), that exhibit a unique ground state, no symmetry 
breaking of any kind, gapped fractionalized ``spinon'' excitations,  
and vanishing correlations in any local order parameter. 
An example of such a spin liquid is realized in the, so called, ring exchange
model 
on the triangular lattice \cite{lhuillier}. }
\item{Resonating Valence Bond spin liquids (Type II), that exhibit no symmetry
    breaking, 
no long range  correlations in any local order parameter, and an extraordinary
    density 
of states in each total $S$ sector.  Numerical work by Dommange et al. 
\cite{Domm03} supports the conjecture of gapless deconfined ``spinon'' excitations in this scenario.
An example of such a spin liquid is believed to be realized by the Heisenberg spin 1/2 model on
the kagom\'e  lattice \cite{lhuillier,subra95,mila98,mambri00,lech97,lechthe95,waldt98,leung93}.}
\end{itemize}

The kagom\'e spin 1/2 antiferromagnet (KAF) seems to be a
paradigmatic example of type II RVB spin liquids, but unfortunately 
so far no experimental realization of this model has been found among solid
state systems.  Only the spin 1 KAF can be  realized in solid state
experiments, 
but that system has a gap 
to all excitations, i.e. it does not belong the type II spin liquids \cite{hita}. 
The physics of the spin $1/2$
KAF is, however, not yet fully 
understood. There are papers that suggest VBC type order with large unit cells \cite{nicolic}.

\subsection{Spinless interacting Fermi gas in a kagom\'e lattice}

We have proposed recently how two realize the trimerized kagom\'e optical
lattices 
using superlattice techniques, and have studied various kinds of quantum gases 
in such lattices \cite{kagome}: 
i) a single component (polarized) Bose gas,
ii) a single component (polarized) interacting Fermi gas,
iii) a two component (``spin'' 1/2) Fermi-Fermi mixture.
In the subsequent paper \cite{kagome2} we have concentrated on the second of
the above mentioned situations
and studied the polarized interacting Fermi gas in the trimerized kagom\'e
lattice at the filling $\nu=2/3$.
Using the method of exact diagonalization of the Hamiltonian we have shown
that the system exhibits novel kind of 
behavior at low temperatures, which has led us to propose a new class of
possible behavior of frustrated AFM's:
\begin{itemize}
\item{Quantum spin-liquid crystal, characterized by the long range N\'eel type of 
ordering at low $T$, gapless spectrum, and anomalously large density of low energy excitations. }
\end{itemize}

This paper is devoted to the presentation of the details of the theory 
described in above mentioned two letters, Refs. \cite{kagome,kagome2}.
First, we discuss briefly the general properties of interactions 
in trimerized kagom\'e lattices as well as the case of a single
component Bose gas in the trimerized kagom\'e lattice. Then we focus, however, 
our attention on a trimerized kagom\'e lattice loaded with a 
spinless Fermi gas with nearest-neighbor interaction. At $2/3$ filling per trimer
such Fermi gas behaves as a frustrated quantum anti-ferromagnet, and exhibits
quantum spin-liquid crystal behavior. The motivation to study this model is
at least fourfold:

i) In a magnetic field such that the trimerized KAF is driven into the
magnetization plateau at 1/3 of the saturation magnetization, the physics
of the KAF is described precisely by our model \cite{subra95,mila98,mambri00}. Studying our model will
thus exactly shed light on the theory of KAF and, hopefully, on
experiments on the KAF.

ii) Theoretical studies (using exact diagonalization of the Hamiltonian) 
indicate that the  model has the fascinating properties of, what we have termed a 
 {\it quantum spin-liquid crystal}.  We expect the behavior observed in this system indeed
to be generic for other ``multi''-merized systems. First of all it is clear 
 that optical methods allow for creating
many similar spin models with couplings depending on bond directions. In
the simplest case this can be accomplished for a square lattice where one
could achieve a ``square lattice of small squares'', for the triangular
lattice to obtain a ``triangular lattice of small triangles'' etc. One can
expect that when such procedures are realized for frustrated systems, this
might lead to similar effects as for the kagom\'e lattice. 

iii) One of the most fascinating possibilities provided by the optical
lattices is the possibility of ``on line'' modifications of the lattice
geometry. We may go from triangular to kagom\'e lattice in real time in a
controlled way. Trimerization (or generally ``multi''-merization) is a new
experimental option, and it is highly desirable to explore its
consequences. Our model (apart from the model of the Bose gas in the
trimerized kagom\'e lattice) is one of the simplest ones to explore these
consequences.

iv) Last, but not least the  model is experimentally feasible.

\subsection{Structure of the paper}

The paper is organized as follows. In section \ref{Sec:Atomic} we briefly describe 
the laser 
arrangement that can be used to create a trimerized optical {\kago} lattice. 
In subsection \ref{Sub:Hubbard}
we first introduce  the Hamiltonian that governs the particle dynamics in the lattice:
it is  a generalized 
Hubbard model that  can be used as a model for bosons, fermions, as well as for 
boson-fermion, or fermion-fermion  
mixtures in the lattice.  We  
show under which conditions a tight-binding description of the particle dynamics is 
appropriate in such a lattice, and present results of the calculations of the Hubbard model couplings
as a function of parameters of the systems and the degree of trimerization. 
In the next subsection (\ref{Sub:Bose_gas_trim}) we present in some detail the
results concerning the physics 
of a Bose gas in the trimerized kagom\'e lattice. Here we generalize the
results   of Ref. \cite{kagome}, obtained in the hard core boson limit, to the 
case when more than one boson can be present at the same lattice site. 
The last subsection (\ref{Sub:Fermi-Fermi}) discusses in short the case of a 
Fermi-Fermi mixture in the trimerized kagom\'e lattice.

In section \ref{Sec:Spinless} we start our discussion of the case of 
$2/3$ filling of the trimerized {\kago} lattice with spinless fermions. We focus our attention 
on the case of strong intra trimer and weak inter trimer coupling. First, we
discuss various methods of creating an ultracold polarized interacting Fermi
gas in an optical lattice (subsection \ref{Sub:Exp}). Then we discuss in detail 
the intra-trimer dynamics.   In the following subsection (subsection
\ref{Sub:Eff}) we show that the low energy physics of such a gas at  $2/3$ filling is described by
an effective spin-$1/2$ Hamiltonian with strongly anisotropic couplings.
The exchange constant $J$ of this Hamiltonian is proportional to the inter trimer atomic interaction 
potential which, in the low-energy limit, can be attractive or repulsive, depending  on the 
species of interacting atoms. In favorable cases it can also 
be manipulated by a magnetic Feshbach resonance. The relation of the model to
the Heisenberg spin 1/2 AFM in the
kagom\'e lattice is discussed in subsection \ref{Sub:Relat}. 
To capture the entire parameter range of the model,
we investigate the properties of the effective spin Hamiltonian for positive and for 
negative exchange coupling. We start our investigations of 
the effective spin model in subsection \ref{Sub:Classic} by looking 
at its classical and semiclassical behavior. 
Surprisingly, we find that for positive 
$J$ there exists a very large manifold of degenerate classical ground states (GS).
The semiclassical spin wave analysis (discussed in the subsection \ref{Sub:SWT} and limited to the most 
symmetric GSs)  
does favor some of those states, but does not give a definite answer 
concerning the real nature of the quantum ground state.  
In section \ref{Sec:Nummer} we present 
the results of exact diagonalizations of finite cells of the realistic spin-$1/2$ version of 
our model. It turns
out that even in this extreme quantum limit the ground state of our model exhibits long-range 
N\'{e}el order of the same structure as is found in the classical version. For positive exchange 
coupling, $J > 0$, we observe a very high density of low-lying eigenstates of the effective 
spin model. We associate these low-lying states  with the manifold of classical ground states 
whose degeneracy is lifted by quantum zero-point fluctuations.
In the concluding section, Sec. \ref{Sec:Disc},
we discuss experimental routes towards verification of our results, 
and detection of the predicted effects.    
The paper contains two appendices, in which we present the details of the calculations of the 
couplings in the Hubbard model, and the mean-field theory of the single component Bose gas, respectively. 

\section{Atomic gases in {\kago}  lattices}
\label{Sec:Atomic}

\subsection{Creation of optical kagom\'e lattices}
\label{Sub:Optical}

In the following, we consider the atoms confined magnetically or optically in the $z$ direction at $z=0$.
The atoms form effectively a  2D system in an optical lattice 
in the $x$-$y$ plane. 
In order to create   a kagom\'e lattice in this plane one can use  red detuned lasers,
so that the potential minima coincide with the laser intensity maxima.
A perfect triangular lattice can be easily created by  
two standing waves on the $x$-$y$ plane, $\cos^2(\k_{1,2}\rm \r)$, 
with ${\bf k}_{1,2}=k \{1/2,\pm \sqrt{3}/2 \}$,
and an additional standing wave 
$\cos^2(\k_3 \r +\phi)$, with $\k_3=k \{0,1 \}$. 
The resulting triangles have a side of length $2\pi/\sqrt{3}k$. 
By varying $\phi$ the third standing wave 
is shifted along the $y$ axis, and, in principle, 
a kagom\'e pattern could be realized. 

Unfortunately, this procedure presents two problems. First, three lasers on a plane cannot 
have  mutually orthogonal polarizations, and consequently 
undesired interferences between different 
standing waves occur. This problem has, however, a relatively simple solution: undesired interferences can be avoided
by randomizing the relative orientation of the polarization between different 
standing waves, or by introducing small frequency mismatches, which, however,
have to be larger than any other relevant frequencies. 
The second problem is much more serious, and is caused by the  diffraction limit. Let us denote 
by $\xi$ the ratio  between the separation 
between maxima of the laser intensity (i.e. minima of the resulting optical potential
in the case of red detuned laser beam) and the 
half-width at half maximum (HWHM). To have a good resolution of the potential minima one needs
 $\xi$ to be definitely significantly
larger than 2. In the case discussed above,
however, $\xi$ is only about $4$ at $\phi=\pi$
in the ideal kagom\'e case. 
Because of that, for any $\phi$, the three potential minima forming the 
kagom\'e triangles cannot be resolved.

\begin{figure}[ht]
\begin{center}
\includegraphics[width=4cm]{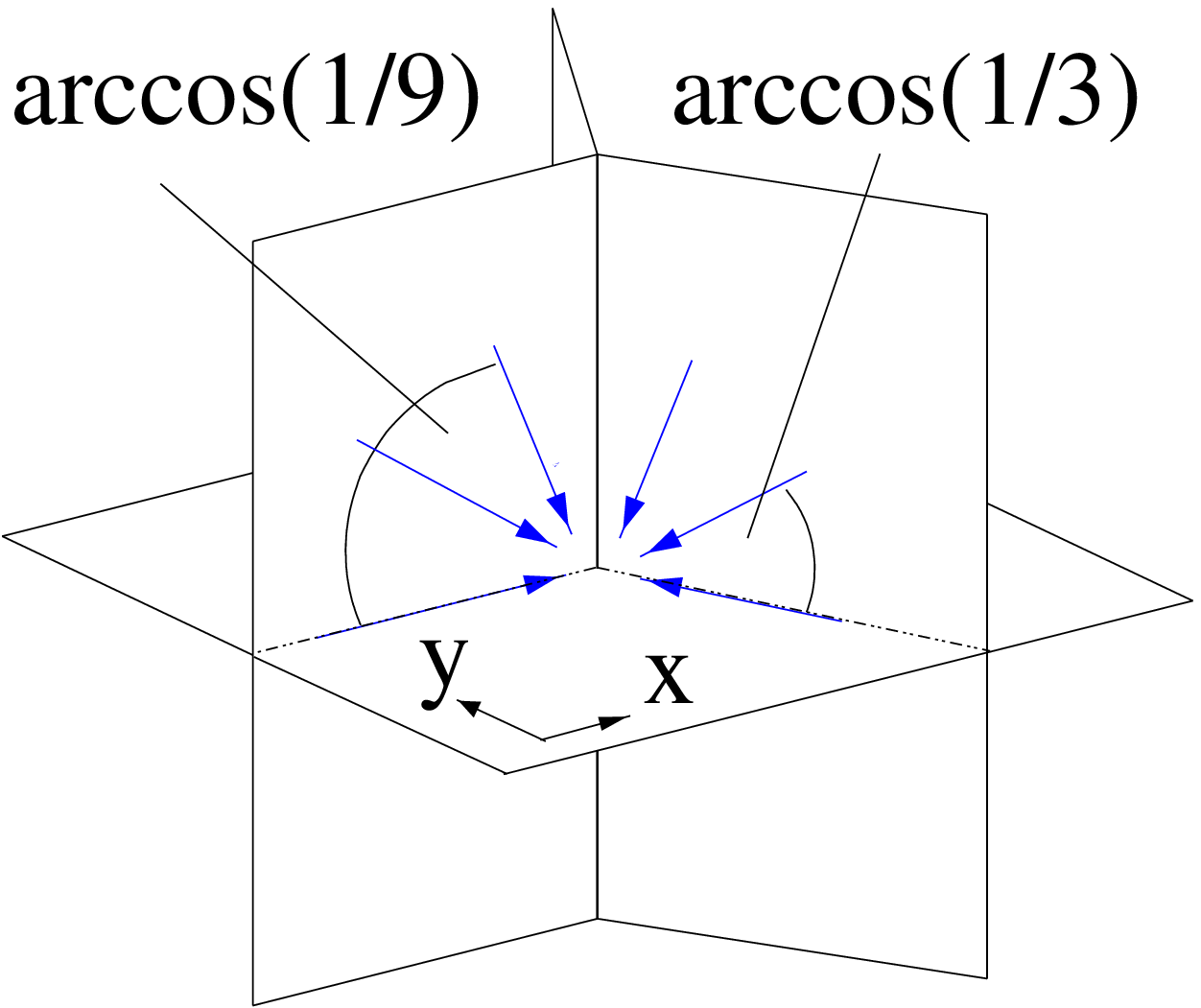}
\includegraphics[width=3cm]{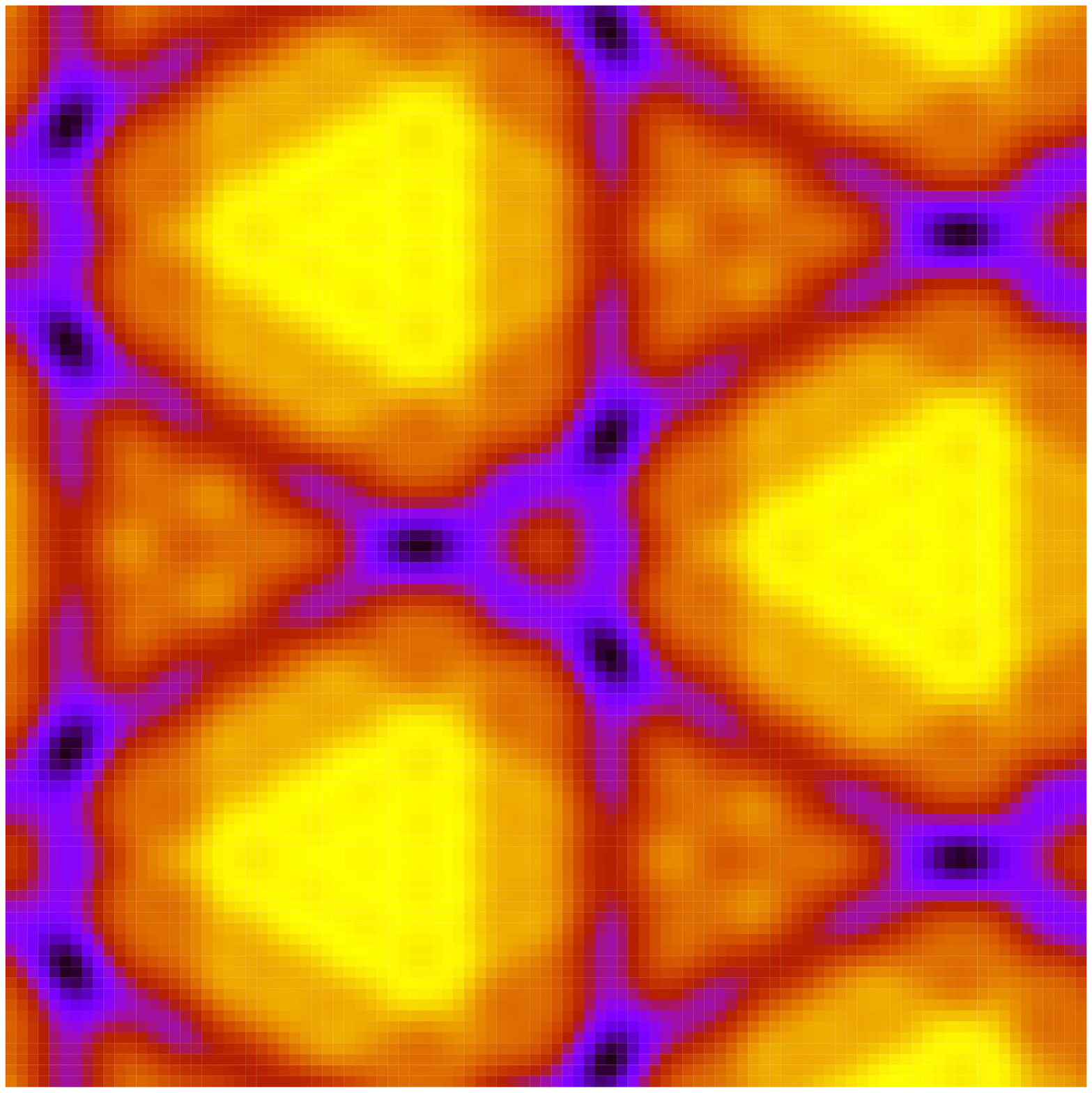}
\end{center}
\caption{(color online) Scheme of the proposed experimental set-up.
Each arrow depicts a wave vector of  a standing wave laser. The three vertical 
planes intersect at an angle of $120^{\circ}$.  
Dark (dark blue in the online version) spots in the 
right kagom\'e figure indicate the potential 
lattice minima.
}
\label{Setup}
\end{figure}

We propose to use the 
super-lattice technique \cite{Superlattice} which we briefly describe
in the following paragraphs, as a method to generate 
ideal and trimerized optical kagom\'{e} lattices. 
The proposed experimental set-up is schematically shown in Fig.~\ref{Setup}. There are three 
planes  of standing-wave laser beams, and the wave vectors of these lasers lie on a plane.
In the particular case of Fig.~\ref{Setup}, we have three standing waves (a
triple) in each plane.  
The laser fields within 
each plane are phase-locked. A  kagom\'{e} lattice will be formed by the intensity 
pattern that results from the sum of the laser intensities of the triples in the $x$-$y$ plane.

In order to resolve the three potential minima in the unit cell of the kagom\'e lattice 
we must use at least two standing waves in each of the three vertical planes shown 
in Fig.~\ref{Setup}.  While the wave-fields in the same plane 
 must have identical polarizations, the fields in different planes should  not interfere.
As mentioned above,  undesired 
interference cross-terms in the total intensity of the fields can be removed either by randomizing the 
relative orientations of the polarizations  between waves from different planes,  or by 
introducing small frequency mismatches.
 With this set-up consisting of 2 waves per vertical plane, we obtain the  
following intensity pattern in the $x$-$y$ plane, 
\bea
I({\mathbf r}) & = & I_0 \sum^3_{i=1}\big[ \cos({\mathbf k}_i {\mathbf r} + \sigma_i \phi/2) 
 \nonumber \\[-3mm] 
&&{}+2\cos({\mathbf k}_i {\mathbf r}/3 + \sigma_i \phi/6) \big]^2,\,\, 
({\mathbf r} = (x,y)),\,\,\,\, 
\label{INTENS} 
\eea
where $\sigma_2 = -1$ and $\sigma_1 = \sigma_3 = 1$ and the index $i$ enumerates the vertical planes. The pattern formed by the maxima of 
the intensity $I({\mathbf r})$ changes between a  triangular lattice at $\phi = 0$,  and 
trimerized {\kago} lattices with varying mesh width for $0 < \phi < \pi$, until at 
$\phi = \pi$ the uniform {\kago} lattice is reached. 
In this   
limit  one obtains the value 
$\xi \approx 7.6$ at $\phi = \pi$. This is sufficient to create a  well resolved ideal 
kagom\'e lattice. Direct inspection shows that in this case a moderately trimerized lattice 
can also be realized : $\xi$ remains sufficiently large for $5\pi/12\le\phi\le\pi$, so that the potential minima can still be resolved.

With the additional third beam shown in Fig.~\ref{Setup}, a resulting intensity pattern
\begin{eqnarray}
        I({\mathbf r}) 
& = & 
        I_0 \sum^3_{i=1}\big[ \cos({\mathbf k}_i {\mathbf r} + 3 \sigma_i \phi/2) 
        \nonumber 
        \\
&+&     
        2\cos({\mathbf k}_i {\mathbf r}/3 + \sigma_i \phi/2) 
        + 4\cos({\mathbf k}_i {\mathbf r}/9 + \sigma_i \phi/6)\big]^2,
        \nonumber
        \\
&&      
        ({\mathbf r} = (x,y)),\,\,\,\, 
\label{INTENS2} 
\end{eqnarray}
is obtained. With this arrangement it
is possible to transform the optical potential smoothly from an 
ideal kagom\'e case into a strongly trimerized lattice. The  value 
of $\xi$ increases in this case to $\approx 14$, and remains large in wide range of angles $\phi$.

\subsection{Hubbard Hamiltonian}
\label{Sub:Hubbard}

Depending on the detuning of the  laser relative to 
the resonance frequency of the atoms, either the minima 
or the maxima of the intensity patterns (\re{INTENS}) and (\re{INTENS2})
form  attractive potentials for the
atoms. If these potentials are sufficiently strong, the tight binding approximation holds \cite{Ashcroft}, and 
 the dynamics of the atomic gas 
can quite  generally be described by a Hubbard type Hamiltonian
\cite{Jaksch, Greiner}:
\begin{eqnarray}
        H_{\rm{Hubbard}} 
&=&
        - \sum_{\langle i j \rangle} t_{ i j } ( c^{\dagger}_i c_j +c_i
          c^{\dagger}_j) + \frac{1}{2} \sum_i U n_i (n_i -1) 
          \nonumber
          \\
&&        
          +
          \frac{1}{2} \sum_{\langle i j \rangle} U_{ij} n_i n_j .
\label{HUBBARD}
\end{eqnarray}
Here  $c^{\dagger}_i$  creates  an atom in a Wannier state localized at the lattice 
site $i$. Depending on the atomic species the operators 
$c^{\dagger}_i$, $c_i$ represent either fermionic or bosonic creation and annihilation 
operators. The parameters $t_{ ij }$, $U$ and $U_{ij}$  of this Hamiltonian are matrix 
elements of the one-particle Hamiltonian and of the interaction potentials of the gas in 
the Wannier representation:
\begin{equation}
t_{ij} = \langle \W_i | H_0| \W_j \rangle,
\label{TUNNEL}
\end{equation}
where $H_0$ is the one-particle Hamiltonian,
\begin{equation}
    H_0 = - \frac{\hbar^2}{2m} \Delta + v({\mathbf r})\,,
    \label{1HAMIL}
\end{equation}
with the one-particle potential $v({\mathbf r}) \propto \pm I({\mathbf r})$--
see (\re{INTENS}) and (\re{INTENS2}). The sign depends on the detuning. 
For the Bose gas interacting via short range Van der Waals 
forces, the scattering at low energies occurs via the $s$-wave channel, 
and is adequately described by the zero-range potential, so that 
\begin{eqnarray}
    U
&=&
    g_{\rm{2D}}
    \int \d^2 x |\W_i(\r)|^4,
    \quad
\label{UONSITE}
\end{eqnarray}
whereas
\begin{eqnarray}
    U_{ij}
&=&
    g_{\rm{2D}}
    \int \d^2 x |\W_i(\r)|^2 |\W_j(\r)|^2 ,
    \quad
\label{UNN}
\end{eqnarray}
where the coupling  $g_{\rm{2D}}=4\pi\hbar^2a_s/m W$ with $m$ the atomic mass, 
and with $W$ the effective transverse width of the 2D lattice in the $z$ direction. 
In the case of polarized fermions $U$ vanishes, since 
$s$-wave scattering is not possible due to the Pauli principle.
The nearest neighbors interaction, on the other hand, are possible, and in the case   
that they are due to dipolar forces (cf.  \cite{Goral}) or similar long range forces the couplings become
\begin{eqnarray}
    U_{ij}
&\sim&
    \int \d^2 x \d^2 x' |\W_i(\r)|^2V(\r-\r\,') |\W_j(\r\,')|^2 ,
    \quad
\label{UNN2}
\end{eqnarray}
where $V(\r)$ is the interparticle potential. 
Obviously, the same expression holds also for bosons interacting via the potential $V(\r)$.
The Hubbard Hamiltonian (\re{HUBBARD}) does not necessarily describe the physics of bare particles;
it may equally well describe the physics of composite objects, such as,
for instance, composite fermions that arise in the analysis of  Fermi-Bose mixtures
in the lattice in the strong interaction limit \cite{Lewenstein}.
The nearest neighbor interactions and tunnelings are  induced by
the original hopping of bare fermions and bosons, and the corresponding
values of $t_{ij}$ and $U_{ij}$ have to be calculated from the
bare couplings following the lines of Ref. \cite{Lewenstein}.  

In this paper we present explicit results  for the tunneling matrix elements $t_{ij}$ and the 
interaction strengths $U$ and $U_{ij}$  in the case of zero-range potential--
expressions (\re{TUNNEL}), (\re{UONSITE}), (\re{UNN}).
To this aim we need to  determine the Wannier functions $\W_i(\r)$ 
for {\kago} type lattices. The method by which this task can be accomplished  is presented 
in details in App.~\ref{App:Param}.

\begin{figure}[ht]
    \includegraphics[width=8.0cm]{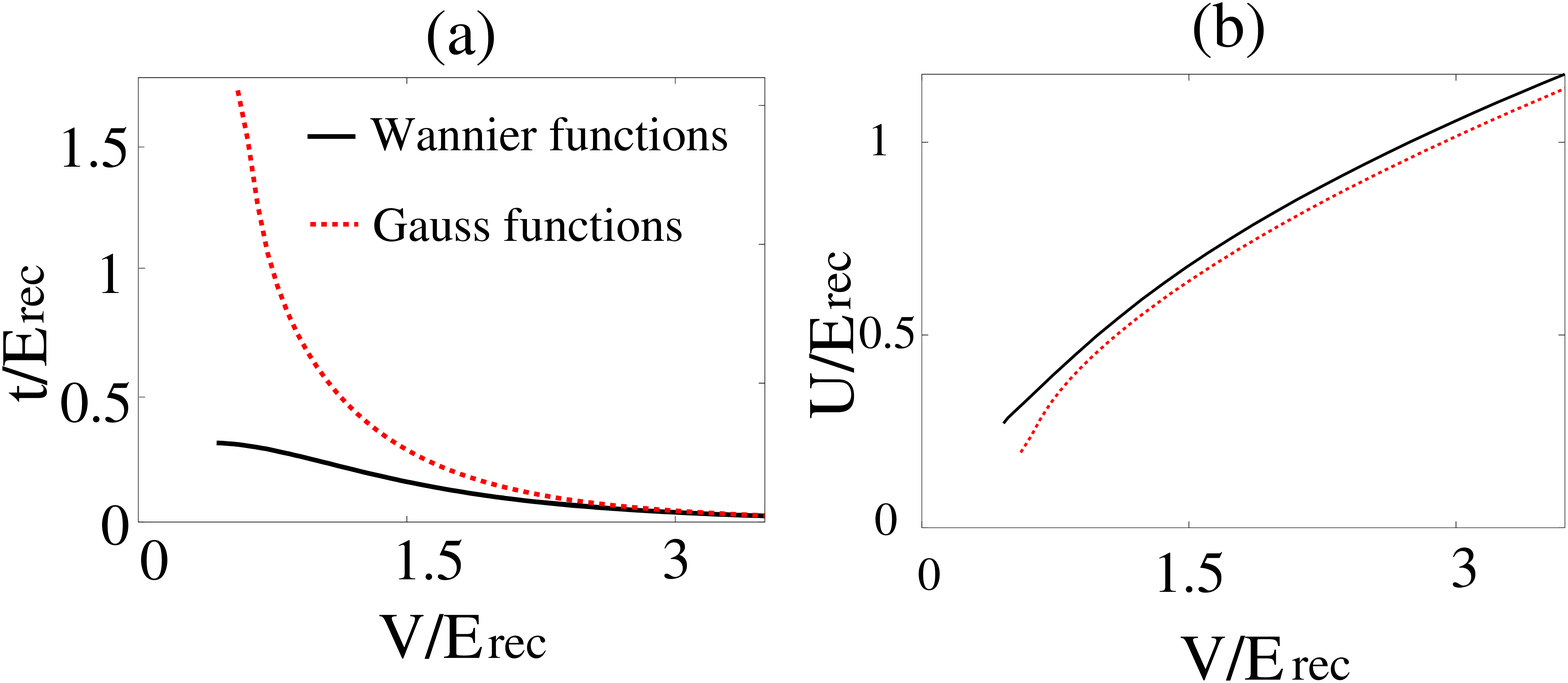}
\caption{(color online) Couplings for a perfect kagom\'e lattice 
obtained using the  Gaussian approximation  (dotted lines)  and using the exact
Wannier functions (solid lines).
Plot (a): hopping matrix elements, plot (b): contact interaction in units of $E_{\rm{rec}}$.
}
\label{fig:gauss_wann}
\end{figure}

For the ideal kagom\'e lattice we have successfully generated the exact Wannier functions,
and calculated the couplings accordingly. These results were then compared 
with the results of the variational method employing a Gaussian ansatz 
(for details see App.~\ref{App:Param}). Fig.~\ref{fig:gauss_wann} compares the results 
calculated with Wannier functions and the Gaussian ansatz.
For moderately strong potentials, say larger than two times the recoil energy $E_{\rm{rec}}$,
  the Gaussian approximation becomes appropriate, 
giving errors less than $50\%$. For sufficiently high potential amplitudes $>5 E_{\rm{rec}}$, the results obtained with the Gaussian approximation
become practically indistinguishable from the exact Wannier results. 

\begin{figure}[ht]
\rotatebox{270}{
\includegraphics[width=3.0cm]{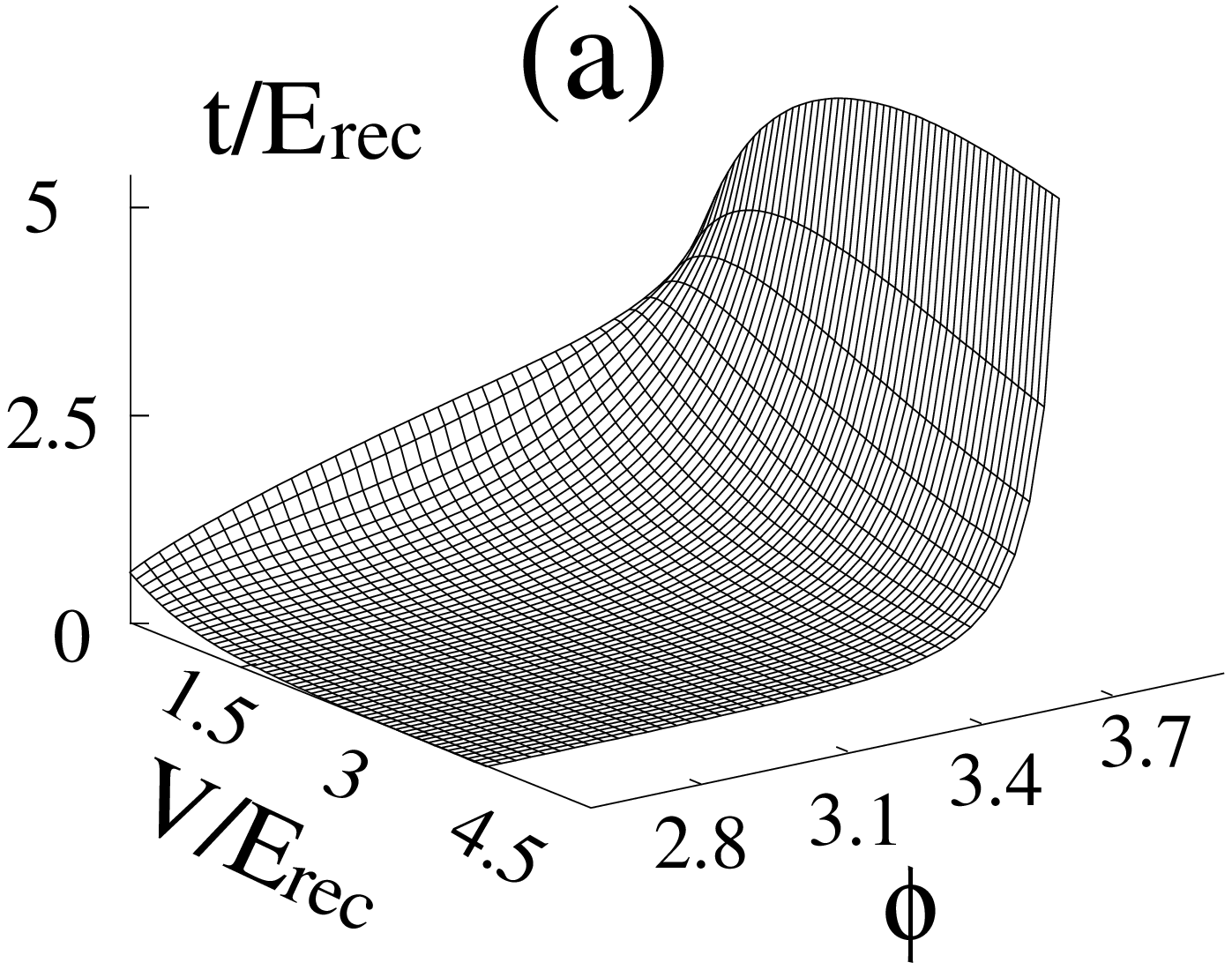}
}
\rotatebox{270}{
\includegraphics[width=3.0cm]{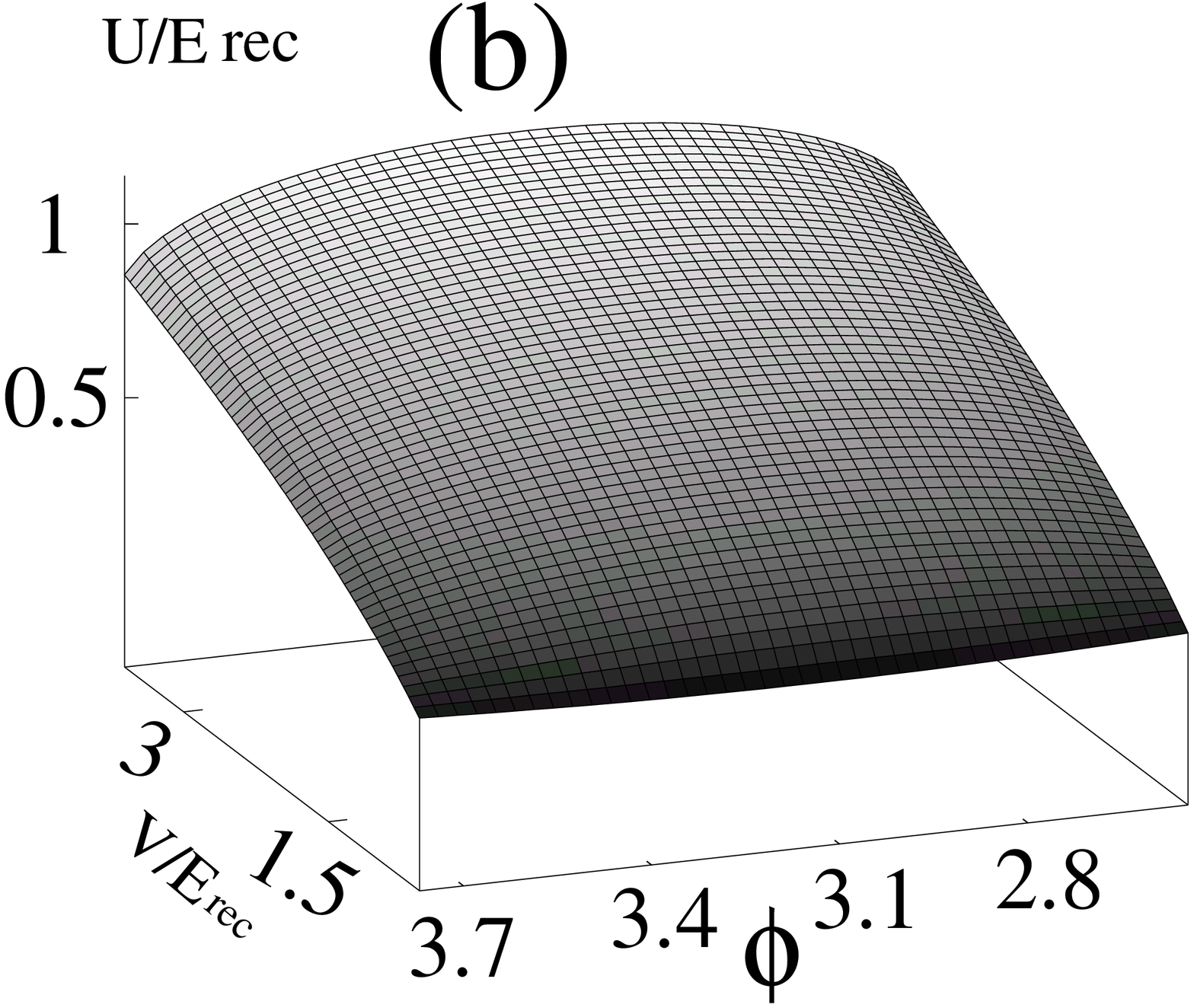}
}
\caption{Couplings for a trimerized lattice obtained with a Gauss-function:
(a) inter-trimer ($\phi\le \pi$) and intra-trimer ($\phi\ge \pi$) hopping matrix elements, 
(b) contact interaction terms in units of $E_{\rm{rec}}$. 
}
\label{fig:gauss}

\end{figure}

Generating well localized Wannier functions in the trimerized lattices is a
difficult task. For this reason,
 guided by the results for the ideal kagom\'e lattice, 
we have limited ourselves here only to the results 
of the Gaussian approximation. Fig.~\ref{fig:gauss} shows the hopping and interaction matrix elements 
depending 
on the trimerization angle. The perfect kagom\'e lattice can be obtained by setting $\phi=\pi$. As 
expected, trimerization does not affect the on-site interactions very strongly, but does change 
the tunneling rates by
orders of magnitude. Already a relatively moderate trimerization introduces large difference between the
inter- and intra-trimer hopping elements.  

\subsection{Bose gas in the trimerized kagom\'e lattice}
\label{Sub:Bose_gas_trim}

As we have mentioned in Sec.~\ref{sec:Introduction}, in the present paper our main focus will be 
on a gas of spinless fermions on the trimerized {\kago} lattice at $2/3$ filling.
Other cases of interest  which we will briefly discuss now include a Bose gas and  a Fermi-Fermi
mixture  in this lattice.

In order to facilitate the calculations, we add to the Hamiltonian
(\re{HUBBARD}) a term of the form 
$-\mu \sum_i n_i$, where $\mu$ is the chemical potential, that controls the average particle number 
of the system. Working with a fixed number of particles is possible, but technically very tedious. 
In the trimerized kagom\'e lattice, the couplings $t_{ij}$ take the values $t$, $t'$ for intra- and inter-trimer hopping, respectively. We set also $U_{ij}=V$ and $=V'$, for intra- and inter-trimer interactions. 

In Ref. \cite{kagome} we have considered the limiting case of hard core
bosons, when $U$ was much larger than any other energy scale, 
i.e. two bosons were not allowed at the same site. 
We have shown then that in  the strongly trimerized case ($ t', V'\ll  V< t$) the system will 
enter a trimerized Mott phase 
with the ground state corresponding to the product over (independent) trimers. 
Depending on the particular value of $\bar\mu\equiv (\mu-V)/(2t+V)$ we may have 
$0$ ($\bar\mu<-1$), $1$ ($-1\le \bar\mu<0$), $2$ ($0\le \bar\mu<1$) or $3$ ($1\le \bar\mu$) 
bosons per trimer, i.e. filling factors $\nu=0$, $1/3$, $2/3$ or $1$ boson per site.
For fractional filling, the atoms within a trimer minimize the 
energy forming a, so-called, W-state \cite{Duer}: 
$|W\rangle=(|001\rangle+ |010\rangle +|100\rangle)/\sqrt{3}$ for $\nu=1/3$, 
and $|W\rangle=(|110\rangle +|101\rangle +|011\rangle)/\sqrt{3}$ for $\nu=2/3$. 
It is worth noticing that $W$-states themselves have interesting applications for 
quantum information theory(c.f. \cite{applW}). 

\begin{figure}[ht]
    \includegraphics[width=8cm]{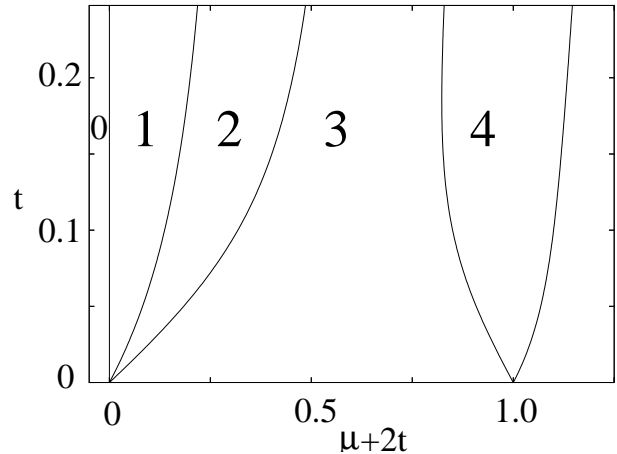}
    \caption{
Mott phases (denoted by the corresponding particle numbers per trimer) 
of the state with lowest energy in the $t-(\mu+2 t)$ plane for 
zero inter-trimer hopping $t'=0$.
}
\label{fig:mott}
\end{figure}
\begin{figure}[ht]
    \includegraphics[width=8cm]{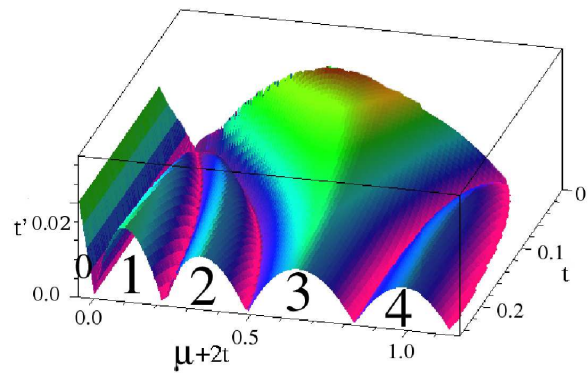}
\caption{(color online) Phase boundaries between Mott- and superfluid-phase in parameter space 
of the hopping elements 
$t$, $t'$ 
and the chemical potential $\mu$.
Below the loops the state is in a Mott-phase, where the number of bosons per
trimer is displayed in the diagram.
}
\label{fig:meanfield}
\end{figure}

Generalizing the Landau mean-field theory
of Ref.~\cite{fisher}, 
we have obtained the phase diagram in the $\bar t'\equiv t'/(2t+V)$ and $\bar\mu$ 
plane
with characteristic lobes describing the boundaries of the Mott phases, given by
$\bar t'=(|\bar\mu|-1)/2$ for $|\mu|\ge 1$, and 
$\bar t'=(3/2)|\bar\mu|(1-|\bar\mu|)/(4-|\bar\mu|)$ for  $|\mu|< 1$. 
Observations of this Mott transition require temperatures $T$ of the 
order of $ t'$, i.e. smaller 
than $ t$ and $ V$. 
Assuming that $U$ is of the order of few recoil energies \cite{Greiner}, 
this  requires  $T$ to be in the range of tens of nK. The results for $ t< V$ are 
qualitatively similar. 

In this paper we present a method to 
generalize these results to the case, when the bosons are not necessarily hard 
core, 
i.e. $U$ may be comparable with $t$.
For simplicity we set $U_{ij}=0$, so that the 
Hamiltonian is still described by the three parameters: $t$, $t'$ 
for intra- and inter-trimer hopping, and $U$ for the on-site interactions. 
Obviously, for vanishing inter-trimer hopping, $t' = 0$, the system is 
in a Mott-insulating state with a fixed number of particles per trimer. 
The corresponding Mott states
are displayed in the phase diagram in the $t-(\mu+2 t)$ plane 
in Fig.~\ref{fig:mott}.   As  $t'$ is 
increased the system undergoes a phase transition into a superfluid state. 
To obtain the phase diagram for this transition, Fig.~\ref{fig:meanfield}, we have 
used a generalization of 
the Landau mean field approach of Fisher et al. \cite{sachdev,fisher,stoof2}, also investigated in
\cite{penna_mean_field}.
Details of the method 
can be found in the Appendix \ref{App:Meanfield}. In our calculations we have confined ourselves 
to values of the chemical potential such that the particle number per trimer does not 
exceed four. In Fig.~\ref{fig:meanfield}, further lobes with higher particle numbers will 
occur along the $\mu$-axes for higher values of $\mu$ than those shown.
Instead of calculating the so called super-fluid order parameter $\psi=\langle b_i \rangle=\langle b_i^\dag \rangle$
self-consistently a fully analytical expressions describing the
boundaries in Fig.~\ref{fig:meanfield} can be obtained, as it is shown in the Appendix \ref{App:Meanfield}.
We mention that by using a cell strong coupling perturbative expansion \cite{penna}
the phase boundary can be obtained with relatively little numerical effort with the accuracy
of a Quantum Monte Carlo simulation.

\subsection{Fermi-Fermi mixture at $1/2$ filling in the trimerized kagom\'e lattice: 
Spin 1/2 Heisenberg antiferromagnet}
\label{Sub:Fermi-Fermi}

For  a fermion-fermion mixture, instead of the extended Hubbard model described by the  Hamiltonian 
(\ref{HUBBARD}) which includes nearest  neighbor interactions $U'_{ij}$, it is more appropriate to 
consider the  Hamiltonian with on-site interactions only:
\begin{equation}
H_{FF} = - \sum_{\langle i j \rangle} t_{ij}(f^{\dagger}_i f_j  
+ \tilde{f}^{\dagger}_i \tilde{f}_j + H.c.) + \sum_i V n_i \tilde{n}_i. 
\label{HFF}
\end{equation}
Here $f_j$ and $\tilde{f}$ denote the fermion annihilation operators for the two species,
and $n_i$, $\tilde{n}_i$ are the corresponding occupation operators. The tunneling  matrix elements 
are $t_{ij} = t$ for intra-trimer and $t_{ij} = t'$ for inter-trimer nearest-neighbor 
tunneling. $H_{FF}$ is then the spin-1/2 Hubbard model. In the strong coupling 
limit, $t,\, t' \ll V$, this model can be transformed into the $t -J$ model 
\cite{auer94} which reduces to the spin-$1/2$ Heisenberg model for half filling,
\bea
H_{FF} \rightarrow H^{Heisenberg} &=& J \sum_{\langle i j \rangle_{intra}} 
{\mathbf S}_i \cdot {\mathbf S}_j \nonumber\\[-3mm]
& & {} + J' \sum_{\langle i j \rangle_{inter}}{\mathbf S}_i \cdot {\mathbf S}_j,\quad 
\label{HHEIS}
\eea
where $J = 4t^2/V$,  $J' = 4t'^2/V$, and ${\mathbf S} = (S^x,\, S^y,\, S^z)$ with 
$n -\tilde{n} = 2 S^z$, $f^{\dagger} \tilde{f} = S^x +\imath S^y$, and 
$\tilde{f}^{\dagger} f = S^x - \imath S^y$. It is exactly the model described by the 
$H^{Heisenberg}$  that has been studied 
by Mila and Mambrini \cite{mila98,mambri00} in their effort to gain a physical understanding 
for the low-lying part of the spectrum of the {\kago} antiferromagnet.  
The physics of this model is very interesting. In the trimerized case, it seems to be clear that 
the system qualifies  as a RVB spin liquid of the second type. The large 
density of singlet and triplet excitations can be predicted quite well by 
analyzing the number of ``relevant'' dimer coverings of the trimerized lattice.
The singlet-triplet gap, if it exists at all, is extremely small.
All of these findings have so far no experimental confirmation.
Experiments on this system are thus highly desirable, and we hope that ultracold atoms will allow 
to realize them. 

\section{ Spinless interacting Fermi gas  in the trimerized {\kago} lattice at $2/3$-filling  }
\label{Sec:Spinless}

\subsection{Experimental realization}
\label{Sub:Exp}

Before we start to discuss the properties of the spinless interacting Fermi gas in the trimerized kagom\'e lattice, we shall first discuss the possibilities of preparing such a system. There are essentially two ways of achieving this goal. First, we may consider
an ultracold gas of fermions that 
interact via dipole-dipole forces.
Bose-Einstein condensation of a dipolar gas of chromium atoms 
has been recently achieved by the group of T. Pfau \cite{pfau}. 
The (magnetic) dipolar interactions in chromium  
are significant, but not very strong. 
There are many ongoing experiments , however, aimed at the creation of ultracold gases of 
heteronuclear molecules, that could carry electric dipole moments of the order of a Debye 
(cf. \cite{tiemann}).   
The observation of physics described in this
paper using heteronuclear molecules with such strong dipoles should be possible already at 
temperatures $T\simeq 100$nK. 

Another possibility of creating an interacting Fermi gas is to use the gas of composite 
fermions that appears in the low temperature behavior of  Fermi-Bose mixtures in the limit of strong 
Bose-Bose and Bose-Fermi interactions. As we have mentioned above the physics of such composite 
fermions is described also by an extended Hubbard model, in which the couplings result from virtual  
tunneling processes involving bare fermions and bosons.  In this case the observation of the low 
temperature physics requires achieving low, but not unrealistic temperatures 
$T\simeq 10-50$nK (c.f. Ref. \cite{fehrmann}). 

The low energy states
may be prepared
by employing adiabatic changes of the
degree of trimerization  of the lattice. For instance, one can start with
a completely trimerized lattice; then the filling $\nu=2/3$ may be achieved 
by starting with $\nu=1$ and by eliminating 1 atom per trimer using, for
instance, laser excitations. One can then increase $t'$ and $U'$ slowly
on a time scale larger  than  $1/J$ ($\simeq$seconds).
Alternatively, one  could start with $\nu\simeq 2/3$ in the moderately
trimerized regime. As in
Ref. \cite{Greiner},  the inhomogeneity of the lattice due to the
trapping potential would then allow to achieve the Mott state with
$\nu=2/3$ per trimer in the center of the trap. Nearly perfect 2/3 filling can be reached
by loading a BEC of molecules formed by 2 fermions into a triangular lattice, generating
a MI state
adiabatically, transforming the lattice to a trimerized kagom\'e one, ``dissociating'' the molecules
by changing the scattering length to negative values,
and
by
finally optically pumping the atoms into a single internal state.
Preparing
$\nu=2/3$ might involve undesired heating (due to optical pumping)
which can be overcome by using laser or phonon cooling afterwards (cf. \cite{daley}).
Note that the imperfections of $\nu$ can be described by a ``$t-J$''-kind of model, and are
of interest themselves.

\subsection{Effective spin model}
\label{Sub:Eff}

\begin{figure}[h]    
    \includegraphics[width=6cm]{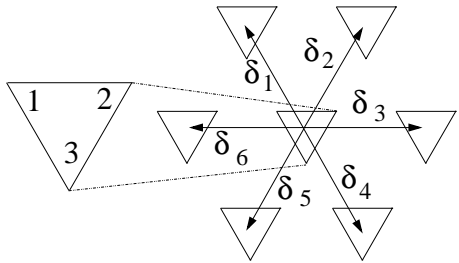}
    \caption{The vectors ${\boldsymbol{\delta}}_i$, $i=1\dots 6$, 
    pointing from the center  of a given 
trimer to the centers of neighboring trimers. Numbering of the sites of the trimers is shown in the 
triangle on the left.}
    \label{star}
\end{figure}

The spinless Fermi gas in the trimerized {\kago} lattice is appropriately
described by the Fermi-Hubbard Hamiltonian
\begin{equation}
H_{FH} = -\sum_{\langle ij \rangle} (t_{ij} f^{\dagger}_i f_j + h.c.) 
+ \sum_{\langle ij \rangle} U_{ij} n_i n_j - \sum_i \mu n_i\,, \label{HFH}
\end{equation}
where $t_{ij}$ and $U_{ij}$ take the values $t$ and $U$ for intra-trimer bonds 
and  $t'$ and $U'$  inter-trimer bonds. $\mu$ is the chemical potential, and 
$n_i = f^{\dagger}_i f_i$ are the occupation numbers with  $f_i$, $f^{\dagger}_i$ 
the fermion annihilation and creation operators. In the following we denote the sites 
of each trimer by $1$, $2$, $3$ in the clockwise sense as shown in Fig.~\ref{star} 

In this section it is our aim to derive from the Hamiltonian (\ref{HFH}) an effective 
spin Hamiltonian that captures in the strongly trimerized limit, $ t',U' \ll t < U$, the low-energy physics the model (\ref{HFH}).\\ 
The intra-trimer part of the Hamiltonian $H_{FH}$ is diagonalized by introducing instead of 
the local fermion  modes $f_1$,  $f_2$ and  $f_3$ the symmetric mode 
$f = (f_1 + f_2 + f_3)/\sqrt{3}$ and the left and right chiral modes 
$f_{\pm} = (f_1 + z_{\pm} f_2 +  z_{\pm}^2 f_3)/\sqrt{3}$ with $z_{\pm}=\exp(\im 2 \pi /3)$:
\begin{equation}
H_{FH}^{intra} = -3tn + t\bar{n} + \frac{U}{2}(\bar{n}^2 - \bar{n}) -\mu\bar{n}\,,
\label{HINTRA}    
\end{equation}
where $n = f^{\dagger}f$ and where $\bar{n} = n +  f^{\dagger}_+  f_+  + f^{\dagger}_- f_-
$ is the total number of fermions in the trimer. In the strongly trimerized limit
the number of fermions is identical in each trimer. It is controlled by the 
chemical potential: for $U + J < \mu < 2U + J$ there are two particles in each trimer, one of them 
occupies the symmetric state, $|1 \rangle = f^{\dagger}|0\rangle$, while the second one occupies either 
one of the chiral states $|1\pm\rangle =  f^{\dagger}_{\pm} |1\rangle$.\\
In the inter-trimer part of the Hamiltonian $H_{HF}$, Eq.~(\re{HFH}), we neglect the hopping term, 
$ -\sum_{\langle \alpha i, \beta j \rangle} t'(f^{\dagger}_{\alpha,i}
f_{\beta,j} + h.c.)$ ($\alpha,\beta = 1,\, 2\,, 3$ refering  to intratrimer
indices and $i$ numbering trimers), 
since any real (first order)hopping process leads to an excited state whose energy is  $\mathcal{O} (U)$ higher 
than the ground-state energy and since second order (virtual) hopping processes yield  
contributions which are small of order $t'^2/U$. Then, the inter-trimer part of $H_{HF}$ 
reduces to 
\begin{eqnarray}
H^{inter}_{HF} &=& \frac{U'}{2} \sum_i ( n_{1,i} n_{3, i+{\boldsymbol\delta}_1} +  
 n_{2,i} n_{3, i+{\boldsymbol\delta}_2} +  n_{2,i} n_{1, i+{\boldsymbol\delta}_3}\nonumber \\ 
 &&
 n_{3,i} n_{1, i+{\boldsymbol\delta}_4} +  n_{3,i} n_{2, i+{\boldsymbol\delta}_5} + 
 n_{1,i} n_{2, i+{\boldsymbol\delta}_6}) 
\label{HINT} 
\end{eqnarray}
Here, $\boldsymbol \delta_{\nu}$,\, $\nu = 1, \, \cdots \, ,6$, denote the six vectors pointing from 
the central triangle to the six neighboring triangles,  see Fig.~\ref{star}.

Next we express the occupation numbers $n_{\alpha,i}$\,, $\alpha = 1,2,3$,\, 
in terms of the fermion 
operators $f$, $f_{\pm}$ (we suppress the site index):
\bea
n_1 & = & \frac{1}{3}\left [\bar n + (f^{\dagger}_+ + f^{\dagger}_-) f + f^{\dagger} (f_+ + f_-)
+ \tau^x \right ]\,,\\
n_2 & = & \frac{1}{3}\left [\bar n + (z_+  f^{\dagger}_+ + z_- f^{\dagger}_-) f + 
f^{\dagger} (z_- f_+ + z_+ f_-)\right.\nonumber\\ &&{}\left. +\cos(2\pi/3) \tau^x + \sin(2\pi/3) \tau^y  \right]\,,\\
n_3 & = & \frac{1}{3}\left [\bar n + (z^2_+  f^{\dagger}_+ + z^2_- f^{\dagger}_-) f + 
f^{\dagger} (z^2_- f_+ + z^2_+ f_-)\right.\nonumber\\ &&{}\left. +\cos(2\pi/3) \tau^x - \sin(2\pi/3) \tau^y  \right]\,.\\
\nonumber
\eea

Here, the (pseudo-)spin operators $\hta^x := \frac{1}{2} (f^{\dagger}_+ f_- + f^{\dagger}_-
f_+)$,\quad $\hta^y := -\frac{i}{2}(f^{\dagger}_+ f_- - f^{\dagger}_- f_+)$ connect the
right- and left-handed chiral fermion states. Inserting  expressions 
($14$) -- ($16$) into $H^{inter}_{HF}$, Eq.~(\re{HINT}),
yields bilinear terms in $\hta^x$, $\hta^y$, linear terms in $\hta^x$ and $\hta^y$, 
bilinear terms in 
$f^{\dagger}$, $f$ and linear terms in $f^{\dagger}$ and $f$. Since none of these terms 
changes the total number of fermions in anyone of the trimers, we may set $\bar n = 2$ in  
the resulting expression for $H^{inter}_{HF}$. However, terms containing the annihilation 
operator $f$ promote the fermion in the symmetric state of a given  trimer into the 
non-occupied  chiral state of the same trimer. A glance at $H_{FH}^{intra}$, Eq.~(\re{HINTRA}), 
shows that the energy of this excited state is ${\mathcal O}(t)$ above the ground-state energy.
Thus, on account of analogous arguments as were given above for the neglect of the 
hopping term of $H^{inter}_{HF}$, we also neglect all terms containing the operators $f^{\dagger}$,
 $f$. The linear terms in $\hta^x_i$, $\hta^y_i$ sum to zero in the sum over the sites $i$ 
so that we arrive at the following effective inter-trimer Hamiltonian 
(we omit an irrelevant constant):
\begin{equation}
H_{eff} = \frac{J}{2} \sum_{i=1}^N \sum_{\nu = 1}^6 \hta_i(\phi_{i,\boldsymbol \delta_{\nu}}) 
\hta_{i+\boldsymbol \delta_{\nu}}(\tilde{\phi}_{i,\boldsymbol \delta_{\nu}})\,. 
\label{HEFF}
\end{equation}
Here, $i$ are the sites of a triangular lattice of $N$ sites on which the trimers are located,
$J = 4 U'/9$, and the vectors $\boldsymbol \delta_{\nu}$, $\nu = 1, \, \cdots \, 6$, are the same as in 
Fig.~\ref{star}. 
In Eq.~(\re{HEFF}), 
$\hta_i(\phi) = \cos(\phi)\hta_i^x + \sin(\phi)\tau^y_i$ and
 $\phi_{i,\boldsymbol \delta_1} = \phi_{i,\boldsymbol \delta_6} = 0$,\quad 
 $\phi_{i,\boldsymbol \delta_2} = \phi_{i,\boldsymbol \delta_3} = 2\pi/3$,\quad  
 $\phi_{i,\boldsymbol \delta_4} = \phi_{i,\boldsymbol \delta_5} = -2\pi/3$,\quad 
 $\tilde{\phi}_{i,\boldsymbol \delta_1} =  \tilde{\phi}_{i,\boldsymbol \delta_2} = -2\pi/3$,\quad  
$\tilde{\phi}_{i,\boldsymbol \delta_3} =  \tilde{\phi}_{i,\boldsymbol \delta_4 } = 0$\quad  and \quad 
$\tilde{\phi}_{i,\boldsymbol \delta_5} =  \tilde{\phi}_{i,\boldsymbol \delta_6 } = 2\pi/3$.

\subsection{Effective spin model: relation to kagom\'e antiferromagnet}
\label{Sub:Relat}

At this point it seems appropriate to briefly discuss the connection between the effective 
Hamiltonian $H_{eff}$ derived here as model for the dynamics of fermionic atoms on a trimerized 
{\kago} lattice and the model Hamiltonian that has been derived by Subrahmanyam \cite{subra95} 
and has later been employed by Mila and Mambrini \cite{mila98,mambri00} to explain the origin 
of the high density of low-lying singlets of  the Heisenberg antiferromagnet (AF) on the {\kago} 
lattice. Mila considers the spin $1/2$ Heisenberg model on the trimerized {\kago} lattice 
with a strong 
intra trimer coupling $J$ and a weak inter trimer coupling $J'$.
In the lowest order 
perturbation expansion with respect to $J'$ he arrives at the effective Hamiltonian
\begin{equation}
H^{trim-kag}_{eff} = \frac{J'}{18} \sum_{\langle ij \rangle} H_{ij}(S_{\bigtriangledown}) H_{ij}(\tau)\,\,,
\label{HMILA}
\end{equation}
where $ H_{ij} (S_{\bigtriangledown}) = \mathbf{S}_{\bigtriangledown i} 
 \mathbf{S}_{\bigtriangledown j}$ and where 
$H_{ij}(\tau)$ is that member of our model $H_{eff}$ that is associated with the bond $ij$.
The operator $\mathbf{S}_{\bigtriangledown i}$ acts on the total spin of the trimer at site $i$,
the trimers form a triangular lattice.
In the derivation $H^{trim-kag}_{eff}$ the Hilbert space of the three $S = 1/2$ spins of the 
individual trimers has been restricted to the subspace of total spin $1/2$ states.  The
four states of this subspace can be be specified by the $z$ - component of their total spin 
and by  two (spin)-chiralities. 
The Heisenberg type Hamiltonian $ H_{ij} (S_{\bigtriangledown})$ acts on the two
 spin states of the trimers 
at sites $i$ and $j$ , $H_{ij}(\tau)$  acts on their chiralities. Obviously, $H^{trim-kag}_{eff}$ 
turns into our model Hamiltonian $H_{eff}$, if the trimer spins $S^z_{\bigtriangledown i}$ are fully 
polarized, e~g~ $S^z_{\bigtriangledown} = 1/2$ for all $i$. This state can be reached by applying a 
sufficiently strong magnetic field to the original trimerized {\kago} AF such that the total 
magnetization reaches $1/3$ of the saturation magnetization, {\it i.~e.~}a magnetic field that 
establishes the $1/3$ magnetization plateau.  

\subsection{Effective spin model: Classical aspects}
\label{Sub:Classic}

As is obvious from the derivation of the Hamiltonian (\re{HEFF}), only its $\tau = 1/2$ 
quantum version can serve as a realistic  effective model for the atomic Fermi gas in the 
trimerized {\kago} lattice.  
Nevertheless, for orientational purposes it is useful 
to first consider this model in the classical limit and to also calculate its 
excitation spectrum in the semiclassical approximation, i.~e.~in the linear 
spin-wave (LSW) approximation. 
We first describe the symmetries of the model  
Eq. (\ref{HEFF}). 

We have found that this model, is not only translationally invariant,
but is also invariant under the point group of order $6$, $Z_6 = Z_3\cdot Z_2$, where the 
generator of $Z_3$ (order $3$) is the combined rotation of the lattice by the angle 
$4\pi/3$ and of the spins by the angle $2\pi/3$ around the $z$ axis, while the 
generator of $Z_2$ (order $2$) is the spin inversion in the $x$-$y$ plane, 
$\hta^x_i \rightarrow -\hta^x_i$, $\hta^y_i \rightarrow -\hta^y_i$. The model 
possesses {\em no} continuous spin rotational symmetry and the lines bisecting the angle 
between two adjacent lattice directions of the triangular lattice are {\em not}
mirror lines.

\begin{figure}
    \includegraphics[width=6cm]{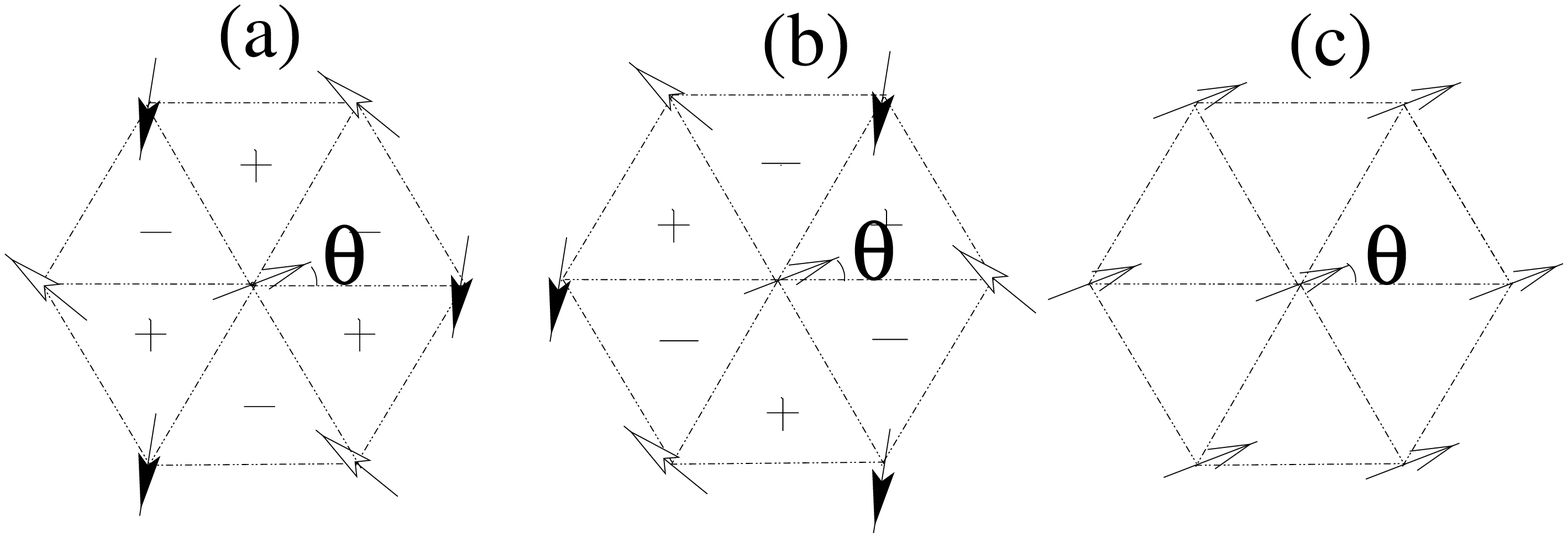}
 \caption{ (a) Classical ground state configuration for $J<0$ (config.~$A$). 
(b), (c) Classical ground state configurations for $J>0$ 
(configuration $B$ and ferromagnetic configuration).
The ``+'', ``-'' signs denote the chirality of the triangular plaquettes (\ref{CHI}). 
 }
 \label{conf7}
\end{figure}

In Figs. \ref{conf7}a, b, c 
we show the three ordered classical states with 
small unit cells on the triangular lattice that are compatible with this point group 
 symmetry: a ferromagnetic state and two 
$120^{\circ}$ N\'{e}el states labeled $A$ and $B$ which differ by the distributions of the 
chiralities $\chi$ over the cells of the lattice as indicated by ``+'' and 
``-'' signs.  
For an elementary cell of the triangular lattice whose corners are labeled $i$,$j$,$k$ 
in the counterclockwise sense, $\chi$ is defined as \cite{miya6} 
\begin{equation}
\chi_{ijk} = (\t^x_i \t^y_j - \t^y_i \t^x_j) + (\quad i,j \rightarrow j,k\quad)
+ (\quad j,k \rightarrow k,i\quad)
\label{CHI}
\end{equation}
$\chi$ is positive (negative) if the spin turns in the counter-clockwise (clockwise) 
sense as one moves around a triangular cell in the counter-clockwise sense. 
Because of the lack of mirror
symmetry mentioned above it is not surprising that the two  
N\'{e}el states have different energies: 
$E^A_{\rm class} = \frac{3}{2}\tau^2JN$,
$E^B_{\rm class} = -\frac{3}{4}\tau^2JN$. 
Here, $N$ is the number of sites and the 
superscripts $A$ and $B$ correspond to the labels of the N\'{e}el states in 
Figs. \ref{conf7}a, b. 

More surprisingly the 
ferromagnetic state is found to be 
degenerate with the N\'{e}el state $B$ in the classical limit, 
$E^{\rm ferro}_{\rm class}= E^B_{\rm class}$. Furthermore, as is indicated by the angle $\theta$ 
in Figs. \ref{conf7}a, b, 
the classical energies of the three  structures 
do not depend on their direction relative to the lattice directions. In summary, in 
the classical approximation the N\'{e}el state $A$ is the ground state (GS) of 
model Eq. (\ref{HEFF}) for negative coupling, $J <0$, while for positive  $J$ there are at 
least two classically degenerate GSs, the N\'{e}el state $B$ and the 
ferromagnetic state. 

\begin{figure}
    \includegraphics[width=6cm]{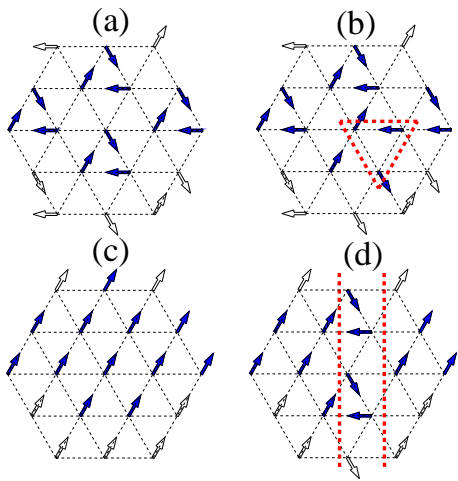}
\caption{(color online) N=12 spin cell: 
(a) Configuration $B$; (b) Localized defect in configuration $B$ marked by the 
triangular contour.
 (c) Ferromagnetic configuration; (d) Line defect in ferromagnetic
 configuration. The open arrows present spins that do not belong to the 12-spin
 cell. Their orientations are determined by the boundary conditions of the cell.
 }
    \label{ground}
\end{figure}

In fact, we have performed a numerical analysis of the 12-spin cell
by fixing the direction of every spin to  $n\pi/3$ with $n=0\cdots5$, so
that there were $6^{12}$ classical spin  configurations. This analysis 
has revealed that for $J<0$ there are $6$ ground states 
each of them exhibiting the N\'{e}el order of type  $A$ (the six fold degeneracy 
comes from a $Z_6$ symmetry of our model). The results are dramatically 
different in the $J>0$ case,  where we have found in 
total $240$ degenerate classical GSs, among which the pure N\'{e}el states of type $B$
and ferromagnetic states sum up to a small fraction. For an illustration,  
see Fig.~\ref{ground} where 
two ordered GS with very large unit cells (Figs. \ref{ground}b, \ref{ground}d) 
together with their parent states  (Figs. \ref{ground}a, \ref{ground}c) are presented.    
As will be seen
below, the large number of degenerate classical GSs may find its analogue in a large 
density of low-lying excitations of the quantum version of  Eq.~(\re{HEFF}).

\subsection{Effective spin model: spin wave theory}
\label{Sub:SWT}

The linear spin-wave expansion around the ferromagnetic GS based on the Holstein-Primakoff 
expansion \cite{auer94} of the spin operators $\hta^x$, $\hta^y$ is straightforward. The spin-wave 
frequency depends on the direction $\theta$  of the magnetization
relative to the main lattice directions of the triangular lattice
 (see Fig.~\ref{conf7}c):
\begin{eqnarray*}
    \of 
&=&
    \frac{3}{2}J\t \sqrt{1 - \frac{4}{3} f({\mathbf q},\theta)}\, 
\end{eqnarray*}   
where 
\begin{eqnarray*}
    f({\mathbf q},\theta)
& = &
    -\sin(\frac{2\pi}{3} + \theta)\, \sin\theta\, \cos({\mathbf q \boldsymbol \delta}_1)
\nonumber\\
                     & &    {}+ \sin(\frac{2\pi}{3} + \theta)\, \sin(\frac{2\pi}{3} - \theta)\,
                            \cos({\mathbf q\boldsymbol \delta}_2) \nonumber\\ 
                     & & {}+ \sin(\frac{2\pi}{3}- \theta)\, \sin\theta \, 
                             \cos({\mathbf q\boldsymbol \delta}_3)\,. \nonumber
\label{oferro}
\end{eqnarray*}

The quantum correction to the GS energy, $\delta E^{\rm ferro}$, is found as
\begin{equation}
\delta E^{\rm ferro}(\theta) = -\frac{3}{4}J\t + 
\frac{1}{2N}\sum_{{\mathbf q}} \of({\mathbf q}, \theta). 
\label{DENFER}
\end{equation}
(In the above expression and till the end of this section 
all expressions for energies  present energies  per site.) 
Evaluating the sum in this expression one finds that $\delta E^{\rm ferro}$ is minimal,
if $\theta$ takes one of the six values $\pi n/3$, $n=0, \cdots , 5$.    
Thus, owing to  
the lowest order quantum corrections to the GS energy the magnetization 
of the ferromagnetic state locks in on one of the directions of the triangular lattice,
{\it i.~e.~} the ferromagnetic GS becomes sixfold degenerate in accordance with the order
of the point group of our model,  Eq.~(\re{HEFF}). With inclusion 
of these lowest order quantum corrections the GS energy for the 
preferred values of $\theta$ is given by
\begin{equation}
E^{\rm ferro} = -\frac{3}{4}J[\t(\t +1) -0.901 \t]
\label{ENFER}
\end{equation}\\
In order to obtain the spin-wave frequencies for the two $120^{\circ}$ structures we 
closely follow the method devised by Jolic{\oe}ur and Le Guillou \cite{joli89} for 
the semiclassical treatment of the Heisenberg AF on the triangular lattice. Since the 
unit cells of both N\'{e}el structures $A$ and $B$ contain three sites one obtains 
 in both cases three surfaces of spin wave frequencies in the magnetic Brillouin 
zone (BZ), $\oa_{\alpha}$ and $\ob_{\alpha}$,\, $\alpha = 1,\, 2,\, 3$ . As for the 
ferromagnetic state, the values of these frequencies depend on the angle $\theta$ 
between these structures and the main  directions of the triangular lattice, and 
hence the quantum corrections to the GS energies of the states $A$ and $B$, $\delta E^A$ 
and $\delta E^A$, depend on $\theta$. For general $\theta$ the expressions for the 
spin-wave frequencies are rather complicated. However, by considering small deviations 
of $\theta$ from the the values $\pi n/3$ we find that as in the ferromagnetic case 
the quantum corrections  $\delta E^A(\theta)$, $\delta E^B(\theta)$ are minimal for 
$ \theta = \pi n/3$, $n = 0 \dots 5$, {\it i.~e.~}the N\'{e}el structures $A$ and $B$ 
also lock in on the directions of the triangular lattice and hence both N\'{e}el states 
are sixfold degenerate. Remarkably, for the structure $B$ all three branches of 
spin-wave frequencies  are dispersionless with the lowest branch consisting of $N/3$ zero 
modes, $\ob_1(\mathbf q) = 0$ for all $\mathbf q$ in the magnetic BZ. This is reminiscent 
of the Heisenberg model on the {\kago} lattice (HAK) for which one also finds $N/3$ 
zero-frequency spin wave modes. The nature of these zero modes is, 
however, quite different for the two models. While they correspond to simultaneous out-of-plane 
rotations  of six-spin clusters in the HAK \cite{harris92}, they represent rigid    
in-plane rotations of the three spins on the corners of an elementary triangle 
in our model (\ref{HEFF}).\\
Since it is of interest, we also note here the expression for the 
GS energy of the state $B$ after the lowest quantum correction has been included:
\begin{equation}
E^B = -\frac{3}{4}J[\t(\t +1) - 1.48 \t]
\label{ENB}
\end{equation}
Comparison of  Eqs.~(\re{ENFER}) and (\re{ENB}) shows that quantum fluctuations lift the 
degeneracy of the purely classical states. In this semiclassical approach it appears 
that the ferromagnetic state is the GS. We recall, however, that there is a 
very large manifold of classical GSs. In this manifold there may well be states which have 
lower energies than the two states that we have compared here.        

\section{Numerical results}
\label{Sec:Nummer}

\subsection{Numerical method}

To describe the physics of spinless fermions on a trimerized optical  
{\kago} lattice at filling $2/3$ we need to consider the model (\ref{HEFF}) 
for spin $\t=1/2$, {\it i.~e.~}in the extreme quantum limit. Questions to be 
answered for this case are: {(\it i)} Is the GS of the model (\ref{HEFF}) an 
ordered state or is it a spin liquid either of type $I$, {\it i.~e.~}a state 
without broken symmetry, with exponentially fast decaying spin-pair 
correlations and a gap to the first excitation, or of type $II$, {\it i.~e.~} 
 a {\kago}-like GS, again without broken symmetry, with extremely short ranged 
correlations, but with a dense spectrum of excitations adjacent to the GS.      
{(\it ii)} What are the thermal properties of our system? After all, the model 
can only be realized at finite, albeit low temperatures.

To find answers to these questions we have performed 
exact diagonalizations (ED) of the the Hamiltonian (\ref{HEFF}) for cells of $N = 12$, 
$15$, $18$, $21$ and $24$ sites using ARPACK routines \cite{arpa}. The sizes of systems
that can be studied by ED are restricted by the amount of memory space that is 
required for storing the non-zero matrix elements. 
To reduce this  requirement we block-diagonalized 
the Hamiltonian (\ref{HEFF}) by exploiting its invariance under $N$-fold translations. 
It allowed us to reduce the problem of diagonalization of $2^N\times2^N$ matrix
to $N$ {\it independent} diagonalizations of matrices of size
$\sim2^N/N\times2^N/N$. This simplification not only lowers the memory
requirements but also greatly reduces the time  of calculation, especially
when a large number of excited eigenstates is of interest.  

Nevertheless, ED of this Hamiltonian remains a demanding task, as in contrast to $SU(2)$ 
invariant spin models the Hilbert space of the Hamiltonian (\ref{HEFF}) cannot be 
separated into subspaces of states with fixed total spin and total $z$-component of the 
spin. 
Because of this last circumstance we had to limit our study to  systems of 
at most $24$ spins. Fortunately, our results for $21$ and $24$ spins show qualitative 
and quantitative resemblance. Therefore we regard them as representative for larger 
systems too. In presenting our results we shall mainly confine ourselves to the two 
largest systems, since  the results for smaller systems suffer from strong finite size 
effects. We remark that only the $12$- and the $21$-site cell can be chosen such that 
these systems possess the full point group symmetry of the infinite lattice. The lack 
of this symmetry for the $15$- and the $18$-site cell adds to the large finite size 
effects observed in the results for these cells.

\subsection{Ground state and low temperature  properties}

\begin{figure}
 \includegraphics[width=7cm]{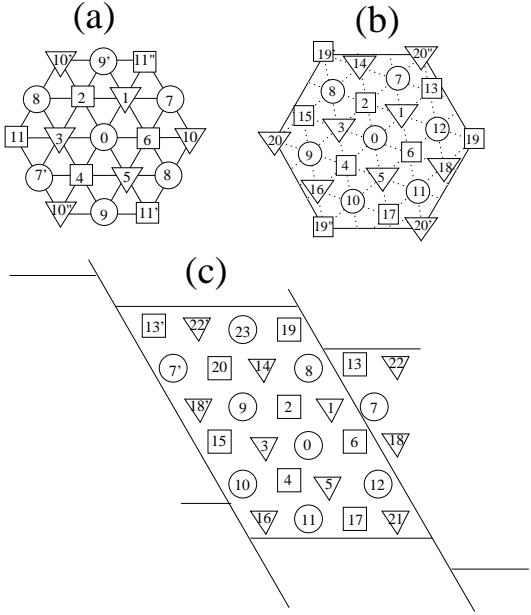}
      \caption{(a) 12 spin cell, (b) 21 spin cell, (c) 24spin cell. 
$\bigcirc$, $\bigtriangledown$, 
and $\Box$ mark the three sublattices.  Primed sites belong to periodic repetitions 
of the cell containing the unprimed sites.}
     \label{cell_all}
\end{figure}

For $J<0$, i. e.  for attractive interaction $U'$ between fermions on nearest neighbor 
trimers, the highest-levels  of $H_{eff}/J$ and the corresponding eigenstates are physically 
most relevant. As will be seen below theses levels are well separated from each other so that 
we only need to calculate a few of them. The situation is drastically different 
in the case $J>0$,  where we need the low-lying states of $H_{eff}$. It turns out 
that there is an abundance of such low lying states. In this respect the spectrum of $H_{eff}$ 
is reminiscent of the spectrum of the Heisenberg Hamiltonian on the 
{\kago} lattice \cite{lech97,waldt98}. 
The answer to the question of whether there is long range order in our model (\ref{HEFF})
is found in Tables \ref{Tab1} and \ref{TabII}, where we show our numerical results for the spatial
spin-spin correlations  for  the $J<0$ and $J>0$ cases, respectively. 
The cells to which these tables refer are shown in Figs. \ref{cell_all}a, b, c. 

Let us first consider the case $J<0$, Table \ref{Tab1}. We have not done a systematic 
finite-size analysis for  these correlations. However, comparing the data for the quantum 
$\t = 1/2$ systems with the classical correlations there can be little doubt that in its GS 
the system orders in the planar $120^{\circ}$ N\'{e}el structure. The smallness of the out-of
plane correlations lends further support to this conclusion.  We have also calculated the 
expectation values of the chirality $ \chi_{ijk}$,\, Eq.~(\ref{CHI}), in the
GSs of the $12$- and of the $21$-site cell and have found perfect agreement
with the pattern of positive and negative chiralities of the classical
configuration, Fig. \ref{conf7}a. Apparently,  for $J < 0$ quantum
fluctuations have a rather weak effect on the GS properties of our model 
(\ref{HEFF}).

\begin{table}[h]
\begin{tabular}{|r||r|r|r|r|}\hline
{} & 1st\! neighbs. & 2nd\! neighbs. & 3rd\! neighbs. & 4th\! neighbs. \\ 
{} & $d=1$        & $d=\sqrt{3}$ & $d=2$ & $d=\sqrt{7}$  \\
\hline \hline   
classical & -0.125 & 0.25 & -0.125 & -0.125 \\
\hline
$N=12$ & -0.137 & 0.251 & -0.125 & {} \\
\hline
$N=21$ & -0.134  & 0.237 & -0.117 & -0.116\\
{}     & (-0.029) & (-0.004) & (-0.004) & (-0.003)\\
\hline
\end{tabular}
\caption{Spin-spin correlations, $ \langle \hat{\tau}^x_0  \hat{\tau}^x_j 
+ \hat{\tau}^y_0 \hat{\tau}^y_j \rangle$ for $J<0$\,. In the last row the 
$\t^z\!-\!\t^z$\, correlations are also shown (numbers in parentheses). 
Owing to the $Z_6$ symmetry of the Hamiltonian (\ref{HEFF}) the correlations 
depend only on the distance $d$ from the central site $0$ (see Fig.~\ref{cell_all}).}
\label{Tab1}
\end{table}

For the case $J>0$ our results for the spin-spin correlations are presented in Table \ref{TabII}.
As for the case $J<0$ we have not been able to perform a finite size analysis but again 
we interpret the data in Table \ref{TabII} as evidence for the existence of planar $120^{\circ}$
 N\'{e}el order in the GS of our model (\ref{HEFF}). This contradicts the prediction 
of the LSW analysis according to which one might have expected to find a ferromagnetic GS
(see subsection \ref{Sub:SWT}). 
However, one must recall that besides the $120^{\circ}$ N\'{e}el GS  and the ferromagnetic 
GS there are many more classical GSs. In a complete LSW analysis one would have to consider 
every one of these states, a task that is practically impossible to perform.  
Relative to the in-plane-correlations the magnitude of  
out-of-plane correlations, which are not displayed in Table \ref{TabII}, 
is even smaller here than
in the case $J<0$. Further support for long-range order in the GS of the model (\ref{HEFF}) 
comes from a comparison of the spin-spin correlations of this model with the same 
correlations of the $=1/2$ Heisenberg AF on the triangular lattice (TAF) which we have included 
in Table \ref{TabII}. It is seen there that the GS correlations of the model (\ref{HEFF}) decay more slowly 
than those of the GS of the TAF which is known to possess  long range $120^{\circ}$  N\'{e}el 
order \cite{bern}. 

Additional strong support for existence of a  N\'{e}el ordered GS in both 
$J>0$ and $J<0$ cases comes from an investigation of chirality patterns.
In both situations the quantum mechanical calculation reveals that there 
exists a perfectly periodic pattern of chiralities $\chi_{ijk}$ 
as in the classical result. 
For $J>0$ we found that  $\chi_{ijk}\approx\mp0.5$ while
for $J<0$ $\chi_{ijk}\approx\pm0.69$. Both results are obtained in $N=21$,
where the $\chi_{ijk}$ was calculated for six triangles located 
around the central site (Fig. \ref{conf7}). The $\mp$, $\pm$ notation indicates 
opposite chiralities 
between  $J>0$ and $J<0$ results. A comparison of these
values to $\mp0.65$ (N\'{e}el B configuration) and $\pm0.65$ 
(N\'{e}el A configuration), leaves little doubt on the nature of 
these GSs. Finally, please notice excellent agreement between quantum and
classical calculation for $J<0$.

\begin{figure}
      \includegraphics[width=8cm]{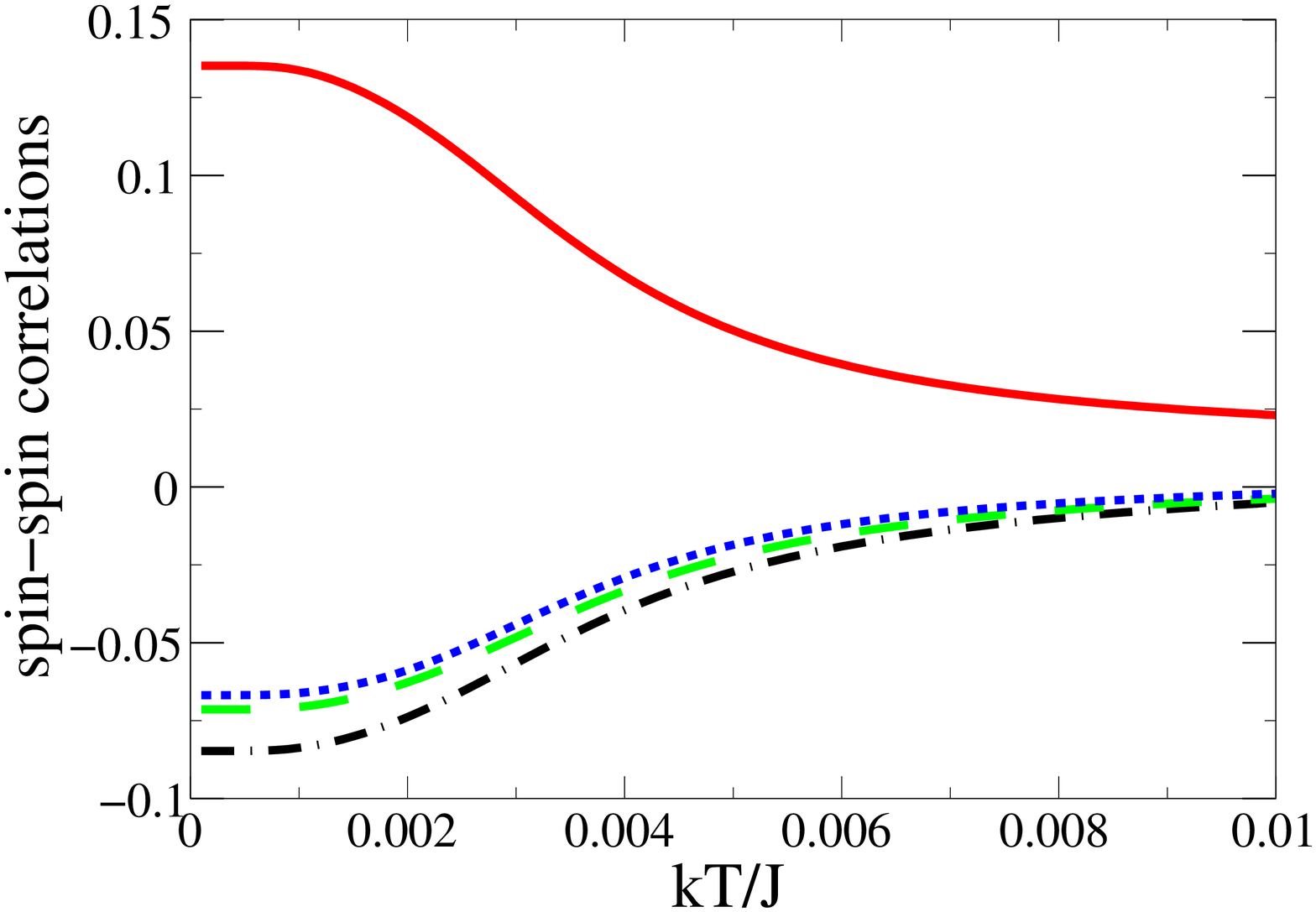}
     \caption{(color online) The curves from 
     a bottom to top correspond to in-plane spin-spin correlation to the 
     1st neighbor (black, dash-dotted), 3rd neighbor (green, dashed), 4th neighbor (blue, dotted) and 
     to the 2nd neighbor (red, solid). Two thousand  lowest eigenstates were used 
     in this calculation. The system size is $N=21$.}
     \label{temperatura}
\end{figure}

Values for the 
spin-spin correlations of the model (\ref{HEFF}) for finite albeit small temperatures are also 
displayed in  Table \ref{TabII}. For $T = 0.005J$ about 800 low lying eigenstates were needed to achieve 
convergence in the data for the correlations. Although these finite 
temperature correlations are smaller in magnitude than the GS correlations, they decay as 
slowly with the distance as the GS correlations {\it i.~e.~} long-range  order persists at 
finite temperatures. This is not surprising since our model (\ref{HEFF}) has {\em no continuous}
symmetry. One thus expects the order to vanish at a finite temperature $T_c$ in a first or 
second order phase transition.
The temperature dependence of spin-spin correlations
in $N=21$ system is depicted in  Fig. \ref{temperatura}.

\begin{figure}
      \includegraphics[width=8cm]{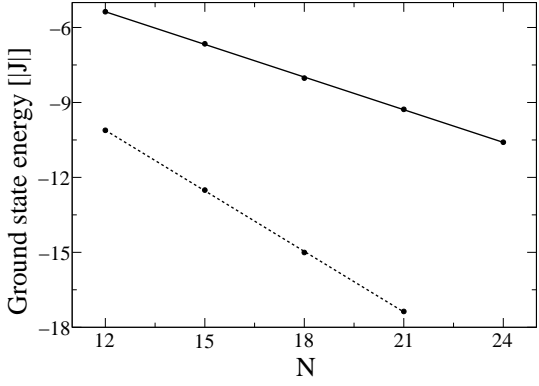}
     \caption{Ground state energies in units of $|J|$ as a function of the
     system size $N$.
     Solid (dashed) line is a linear fit to $J>0$ ($J<0$) data.
     The fit gives in $J>0$ case: $ -0.22 \, J \, N - 0.07 \, J$, while in the 
     $J<0$ case: $0.40 \, J \, N + 0.20 \, J$.
     }
     \label{e0}
\end{figure}

The finite size effects affect the correlations very 
strongly for system sizes $N<21$. 
In Fig. \re{fig:honecker} we plot the spin-spin correlations for 
the various system sizes. The data for N=15, 18 are averages of the 
raw data for fixed lattice distances over the lattice directions. Because of 
boundary effects the correlations do not show the expected six fold symmetry. 
As a consequence the data for N=15, 18 cannot be used in a finite-size 
extrapolation. 
Nevertheless, despite these large finite size effects, for both cases $J>0$ and $J<0$ the GS energies can be reliably extracted from
the data for all the  cell sizes including the smaller ones. From the linear fits shown in 
Figs.~\ref{e0}a, b we obtain $E^A_{GS} = -0.40|J|$ as the GS energy in case $A$. This 
is to be compared with the classical GS energy (see subsection \ref{Sub:Classic}): 
$E^A_{\rm class} = - \frac{3}{2}\t^2|J| = -0.375|J|,\,\, (\t = 1/2)$. In the same way we find 
$E^B_{GS} = - 0.22J$ as the GS energy per site in case $B$, which is to be compared to the 
classical GS energy (see subsection \ref{Sub:Classic}) 
$ E^B_{\rm class} = - \frac{3}{4}\t^2J = -0.1875J$.

\begin{figure}
      \includegraphics[width=8cm]{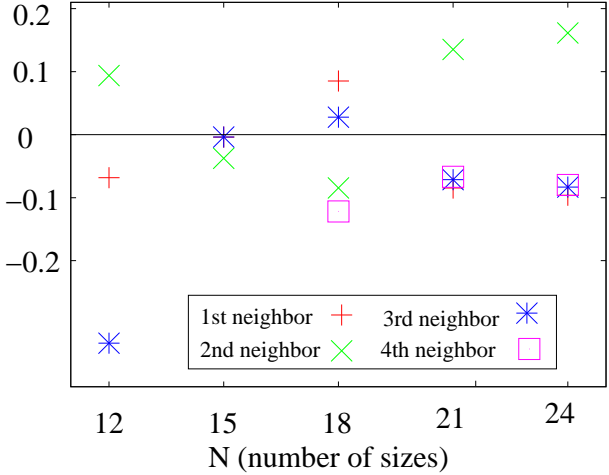}
     \caption{(color online) Spin-spin correlations for 
the various system sizes. The data for N=15, 18 are averages of the 
raw data for fixed lattice distances over the lattice directions.}
     \label{fig:honecker}
\end{figure}

\begin{figure}
      \includegraphics[width=7cm]{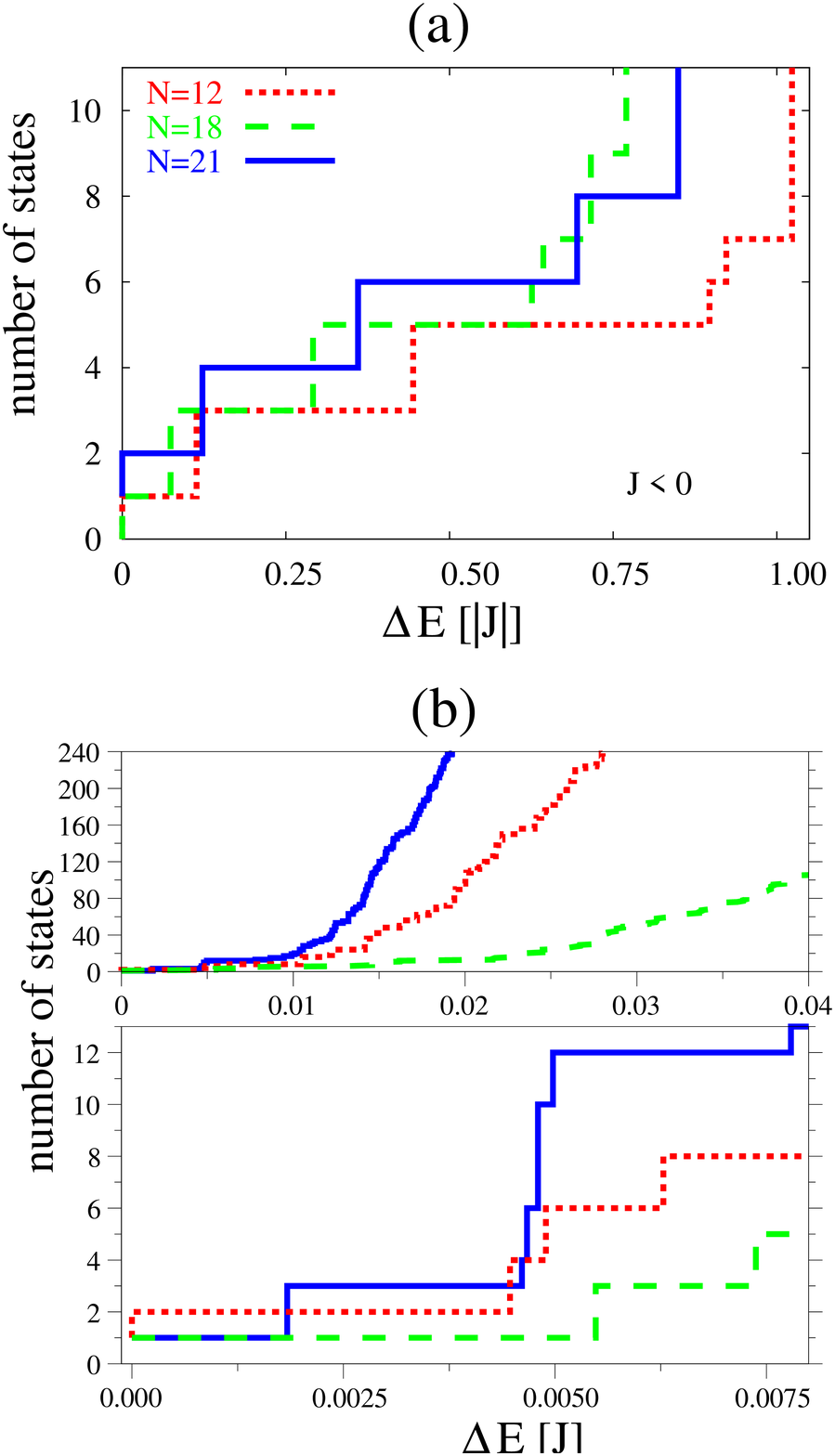}
     \caption{(color online) Accumulated density of states, {i.~e.~}
     the number of states in the 
      energy interval $\triangle E$ above the ground state for (a) $J<0$ and (b) $J>0$.
      The curves in plot (b): blue (solid) N=24, red (dotted) N=21 and green
      (dashed) N=18.}
     \label{stdenJ}
\end{figure}

\subsection{Low energy spectra}

Let us finally discuss the energy spectra of our model for both cases $J<0$ (A), and $J>0$ (B). 
Figs.~\ref{stdenJ}a, b show the accumulated density of states (accumulated DOS) of 
our model (\ref{HEFF})
for the two cases. On account of the breaking of the discrete symmetries of the Hamiltonian 
(\ref{HEFF}) by the $120^{\circ}$ Ne\'{e}l order the standard expectation would be that the GS
is sixfold degenerate for the {\em infinite} model and, since there is no continuous 
symmetry that could be broken, the excitations should be separated from the GS by a finite gap 
of the order of $J$. For {\it finite} systems the GS degeneracy will be lifted. Nevertheless,
we expected to find six low-lying states in the gap below the lowest excited state.
Fig.~\ref{stdenJ}a, $J<0$,  does not reflect this scenario convincingly.  However, there are 
only a few states with energies substantially below $0.5|J|$. We take this as an indication   
of a gap of this order of magnitude in the spectrum of the Hamiltonian (\ref{HEFF}) in the 
thermodynamic limit $N \rightarrow \infty$.

Obviously, for $J>0$ the spectrum differs drastically from the above expectations, see 
Fig.  \ref{stdenJ}b. There is an abundance of very low-lying excitations, e. g. for $N=21$ 
there are about $2000$ ($800$) states with energies less than $0.09J$ ($0.05J$)
above the ground state.
From the  perfect symmetry  of a finite temperature spin-spin correlations
and their relatively slow decay with temperature, 
we conclude that majority of these excited eigenstates support the spin order of the GS.

Comparison of the lower panel of Fig.~\ref{stdenJ}b
with the scenario outlined above suggests that the gap, if any, is smaller than 
$0.5\cdot 10^{-2}J$. 

\begin{figure}
      \includegraphics[width=7cm]{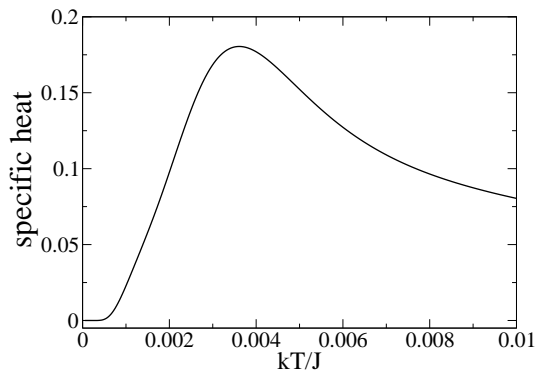}
     \caption{Specific heat for the $N = 21$-- Eq. (\ref{sh}). Two thousand  lowest eigenstates 
     were used in the formula (\ref{sh}) to prepare this plot. 
     }
     \label{cv}
\end{figure}

The rapid increase of the accumulated DOS that sets in at excitation 
energies of this order of magnitude leads to  peaks  in the specific heat,
\begin{equation}
\label{sh}
\frac{1}{N}\frac{\partial }{\partial T} \frac{\sum_i E_i \exp[-E_i/(kT)]}{\sum_i \exp[-E_i/(kT)]},
\end{equation}
at 
the corresponding temperature. We have checked that the peak shifts towards lower temperatures
as the size of the system increases. 
Indeed, in the $N= 24$ system, we found the  peak at 
$kT\approx2.5\cdot10^{-3}J$ while for $N=21$ it is at 
$kT\approx3.6\cdot10^{-3}J$,  see Fig. \ref{cv} for the $N=21$ results. 
The precise determination of the peak  position 
and amplitude in the $N=24$ system requires, 
however, a calculation based on more excited eigenstates
than we have been able to get ($\sim240$, see Fig. \ref{stdenJ}b). 
The shift of the peak position between these two systems reflects the slightly different 
behavior of their accumulated DOS. It would be very interesting to know, 
whether this trend continues for larger $N$ so that in the thermodynamic 
limit, $N \rightarrow \infty$, the specific heat no longer decreases to zero at $T = 0$.
In this case our model (\ref{HEFF}) would be an example of a quantum model with a finite 
zero-temperature entropy.  

Because of the strong finite size effects in the data for $N < 21$ the growth law 
of  the accumulated DOS with the system size $N$ cannot be extracted reliably from 
our data. They are compatible, however, with an exponential increase of the number 
of low-lying states with $N$. 


The features of the low-energy part of 
the spectrum of our model, Eq. (\ref{HEFF}),
are strongly reminiscent of the low-energy part of the  spin-$1/2$ Heisenberg 
antiferromagnet on the {\kago} lattice (HAK) \cite{lech97, waldt98}. There is, however, one  
decisive difference between the two models: while all GS correlations were found to be 
extremely short ranged \cite{leung93} in the HAK there is, in all probability, long range 
spin order in the GS of our model. The absence of long range order in the HAK led 
Mila and Mambrini \cite{mila98,mambri00} to study the trimerized HAK in the basis consisting 
of  all independent dimer coverings of the lattice by exclusively nearest-neighbor singlet 
pairs. By definition this restricted basis cannot produce any long range order in the GS 
of the HAK. Using it in analytical and in numerical calculations Mila and Mambrini were able 
to reproduce the low-lying part of the spectrum of the HAK. In particular, they were able 
to determine the constant $\alpha$ 
in the growth law $\alpha^N$ that describes the increase of the number of low-lying states 
in the HAK. However, the approach of Mila and Mambrini is not suited for the treatment of  our 
model for at least two reasons: {\it(i)} in contrast to the HAK our model is not $SU(2)$ 
invariant. Therefore a restriction of the full Hilbert space of the model to exclusively 
singlet states is unwarranted. {\it(ii)} We need to describe spin-ordered states, and this is
not possible in a basis consisting of products of nearest-neighbor singlet pairs.
 We suggest that for our model (\ref{HEFF}) the abundance of 
low-lying quantum states corresponds to the abundance of classical GS described in 
subsection \ref{Sub:Classic}. 
Zero-point fluctuations lift the degeneracy of the classical 
states leaving the spin correlations that are built into these classical states qualitatively 
untouched. On account of its low-energy properties we have proposed the name 
{\it quantum spin-liquid crystal} for our system.

  \begin{table}[h]
\begin{tabular}{|r||r|r|r|r|r|}\hline 
j & classical & triang. &\multicolumn{2}{|c|}{$N=21$} &$N=24$\qquad\\
 &  & ~Heis. AF &\multicolumn{2}{|c|}{}&\\
\hline
 {} & $T=0$ &  $T=0$ & $T = 0$ & $T = 0.005 J$ &  $T = 0$\\
\hline
$1$st\! nghbs. & {} & {} & {} & {} & {} \\
1 & -0.125         & -0.125        & -0.0847       & -0.027        & -0.0964\\ 
2 & $\cdot \quad$  & $\cdot \quad$ & $\cdot \quad$ & $\cdot \quad$ & -0.0957\\
3 &  $\cdot \quad$ &   $\cdot \quad$ & $\cdot \quad$ & $\cdot \quad$ & -0.0964\\
4 &  $\cdot \quad$ & $\cdot \quad$ &  $\cdot \quad$ &  $\cdot \quad$  & -0.0964\\
5 &  $\cdot \quad$ & $\cdot \quad$ & $\cdot \quad$ &   $\cdot \quad$ & -0.0957\\
6 &  $\cdot \quad$ & $\cdot \quad$ & $\cdot \quad$ &   $\cdot \quad$ &  -0.0964\\
\hline
2nd\! nghbs.  & {} & {} & {} & {} & {}\\
7 &  0.25 & 0.102  & 0.1352 & 0.050 & 0.163\\
8 &  $\cdot \quad$ & $\cdot \quad$ & $\cdot \quad$ &   $\cdot \quad$ & 0.1605\\
9 &  $\cdot \quad$ & $\cdot \quad$ & $\cdot \quad$ & $\cdot \quad$ & 0.1605\\
10 &  $\cdot \quad$ & $\cdot \quad$ & $\cdot \quad$ & $\cdot \quad$ & 0.1630\\
11 &  $\cdot \quad$ & $\cdot \quad$ & $\cdot \quad$ & $\cdot \quad$ & 0.1605\\
12 &  $\cdot \quad$ & $\cdot \quad$ & $\cdot \quad$ & $\cdot \quad$ & 0.1605\\
\hline 
3rd\! nghbs. & {} & {} & {} & {} & {}\\
13 & -0.125 & -0.037 & -0.0714 & -0.022                             & -0.0830\\
14 &  $\cdot \quad$ & $\cdot \quad$ & $\cdot \quad$ & $\cdot \quad$ & -0.0833\\
15 &  $\cdot \quad$ & $\cdot \quad$ & $\cdot \quad$ & $\cdot \quad$ & -0.0830\\
16 &  $\cdot \quad$ & $\cdot \quad$ & $\cdot \quad$ & $\cdot \quad$ & -0.0830\\
17 &  $\cdot \quad$ & $\cdot \quad$ & $\cdot \quad$ & $\cdot \quad$ & -0.0833\\
18 &  $\cdot \quad$ & $\cdot \quad$ & $\cdot \quad$ & $\cdot \quad$ & -0.0830\\       
\hline
4th\! nghbs.  & {} & {} & {} & {} & {}\\
19 & -0.125        & -0.044        & -0.0668       & -0.019        & -0.0799\\
20 & $\cdot \quad$ & $\cdot \quad$ & $\cdot \quad$ & $\cdot \quad$ & -0.0799\\
21 & $\cdot \quad$ & $\cdot \quad$ & $\cdot \quad$ & $\cdot \quad$ & -0.0799\\
22 & $\cdot \quad$ & $\cdot \quad$ & {} & {} & -0.0799\\
\hline
5th\! nghb.  & {} & {} & {} & {} &\\ 
23 & 0.25 & 0.076 & {} & {} & 0.1563\\
\hline
\end{tabular} 
\caption{ Spin-spin correlations as in Table I, but for $J>0$. Sites $j$ are 
numbered as in Fig.~\ref{cell_all}. For comparison  
the correlations $\langle S^x_0 S^x_j +  S^y_0 S^y_j \rangle$ for the spin-$1/2$ Heisenberg AF 
on the triangular lattice are also displayed (data from \cite{lechthe95}). Dots below a value 
for the correlation indicate that this value occurs repeatedly.}
\label{TabII}
\end{table}     

\section{Conclusions}
\label{Sec:Disc}

In this paper we have discussed in detail the physics of ultracold
gases in trimerized kagom\'e lattices. Observation of this kind of physics, 
and detection of the predicted effects requires various steps:
preparation of the trimerized lattice, loading of 
the considered gases, and 
detection. The first step, i.e. the preparation of the kagom\'e 
lattice, is discussed in detail in subsection \ref{Sub:Optical}. 

Probably, the easiest experiment to perform concerns the observation 
of the novel Mott phases of the Bose gas. 
Temperature requirements ($\simeq 100$nK) are rather moderate. 
The challenging problem here is how to achieve $1/3$, $2/3$ fillings, etc. 
In principle physics should do it for us, since the ``exotic'' Mott
phases are the thermodynamic phases of the system at zero temperature. 
There is, however, another elegant method of preparing such phases. 
To this aim one should start with a triangular lattice and achieve a Mott
state with 1, 2, 3, ... atoms per site. Then one should deform the lattice to
a trimerized kagom\'e. The detection of such Mott phases can be done simply 
by releasing the atoms from the lattice, as in Ref. \cite{Greiner}. 
Coherence on the trimer level will then be visible in the appearance of
interference fringes in time-of-flight images, which should reflect the
on-trimer momentum distribution $\sim \sum_{i,j} \cos \vec k (\vec r_i-\vec r_j)$,
where $\vec r_i$ are the positions of the minima in a trimer. In spite of the
appearance of these fringes, the Mott-insulator nature of the state would be 
apparent in the presence of a gap for the excitations, which can be observed by
tilting experiments as those of Ref.~\cite{Greiner}. The opening of the gap
should be analyzed as a function of the trimerization degree $t'/t$, which can be
controlled as discussed in Sec.~\ref{Sub:Optical}.

The experiment with Fermi-Fermi mixture is more demanding. The main problem is, of course, the preparation
of the states in the low energy singlet sector. One  possible way to  prepare  a singlet state in the 
trimerized kagom\'e lattice
with $T<3t/4$ could employ  the recently obtained 
Bose-Einstein condensates of molecules consisting of two fermionic 
atoms \cite{jin} at temperatures of 
the order of $10$ nK. Such BECs should be loaded onto an ideal and 
weak kagom\'e lattice. 
Note that the molecules formed after sweeping 
across a Feshbach resonance, are in a singlet state of the 
pseudo-spin $\vec s$.
This can easily be seen, because the two fermions enter the resonance 
from the $s$-wave scattering channel (i.e. in the symmetric state with 
respect to the spatial coordinates), and thus are  
 in a singlet state of the pseudo-spin (i.e. antisymmetric state with 
respect to exchange of electronic and nuclear spins). 
Since the interaction leading to the 
spin flipping at the Feshbach resonance \cite{Timmermans} 
is symmetric under the simultaneous interchange of both 
nuclear and electronic spin, then the formed molecule 
remains in a pseudo-spin singlet. The typical size of the 
molecule is of the order of the $s$-wave scattering length $a$, 
and thus can be modified at the resonance \cite{Petrov},
being chosen comparable to the lattice wavelength. 
Growing the lattice breaks the molecule into two separate fermionic 
atoms in neighboring sites  in the singlet pseudo-spin state. 
In this way, a singlet covering of the kagom\'e lattice may be achieved,
allowing for the direct generation of a RVB state \cite{anderson}.

Assuming that we can prepare the system in a singlet state at $J'<T<J$, 
then the density of states of the low lying singlet levels can be obtained 
by repeated measurements of the system energy. 
The latter can be achieved by simply releasing the lattice, so that after 
taking care of the zero point energy, 
all of the interaction energy is transformed into kinetic energy. 
In a similar way we can measure the mean value and the distribution 
of any nearest-neighbor two-spin correlation functions. To this aim one has to  
 apply at the moment of the trap release a chosen nearest neighbor two-spin 
Hamiltonian and keep it acting during the cloud expansion 
(for details see \cite{spins}).  In a similar manner we can measure the 
spectrum of triplet excitation, by exciting a triplet state, 
which can be done by flipping one spin 
using a combination of superlattice methods and laser 
excitation \cite{blochpri}. The measurement of the singlet-triplet gap
requires a resolution better than $J'$.

A similar type of measurements can be 
performed in the ideal kagom\'e lattice, when $J=J'$. In this case, 
the singlet-triplet gap is filled with singlet excitations \cite{waldt98}. 
By varying $\phi$, one can 
transform adiabatically from strongly trimerized to ideal Kagom\'e, for which 
the final value of $J$ will be smaller than the initial $J$, but larger than the initial $J'$. 
In that case, the system should remain within the lowest set of $1.15^N$ states
that originally formed the lowest singlet band. The singlet-triplet 
gap, if any,  is estimated to be $\le J/20$, and should be measurable 
using the methods described above. 

The observation of properties of the spinless interacting Fermi
gas is also experimentally very challenging. The first step is to create the interacting Fermi gas, 
obviously. As we discussed it in subsection \ref{Sub:Exp}
this can be achieved either with dipolar particles, or composite fermions.
Both of these possibilities are challenging themselves, although the rapid progress in 
cooling and trapping of dipolar atoms \cite{pfau} and molecules 
allows one to hope that interacting spinless Fermi gases will be routinely available in the next future. 
Preparing of the 2/3 filling is also a challenge, but several 
routes have been proposed in subsection \ref{Sub:Exp}.
Yet another challenge is to measure the predicted properties of the quantum spin-liquid crystal. 

One quantity which should be possible to measure relatively easy, 
is the energy of the system. This can be done 
simply by opening the lattice; by repeated
measurement of the energy $E(T)$ at (definite) finite temperatures one would
 get in this way an
access to the density of modes, i.e. one  could compare the results with Fig.
\ref{stdenJ}. From such measurements one could infer
the existence of a gap $E_{\rm gap}$, since, if  $E_{\rm gap}$
 is large enough,  $E(T)$ becomes $T$-independent for $kT\le E_{\rm gap}$.
Various other
correlations could be measured using the
methods  proposed in Ref. \cite{spins}. In order to measure planar spin correlations,
one has, however, to lift
the degeneracy of the $f_{\pm}$ modes, e. g. by slightly modifying the
intensity of one of the superlattices forming the trimerized lattice.
This should be done on a time scale faster than the characteristic time
scales of other interactions, so that the state of the system would not
change during the measurement.  In such a case one can use far off
resonant Raman scattering (or scattering of matter waves) to measure the
dynamic structure factor, which is proportional to the spatio-temporal
Fourier transform of the density-density correlations. At frequencies
close to the two photon Raman resonance between the
$f_{\pm}$ modes, only $f_+$--$f_-$ transitions contribute to the
signal, and hence such measurement yields the desired information about
the correlations
$\langle f_+^{(i)\dag}f_-^{(i)}f^{(j)\dag}_-f_+^{(j)}\rangle$,
and the spin correlations of Figs. \ref{temperatura},\ref{fig:honecker}. 

We acknowledge discussions with M. Greiner, R. Grimm and P. Julienne. 
We are especially indepted to A. Honecker for his collaboration 
on this project. This paper has been supported
by the Deutsche
Forschungsgemeinschaft (SFB 407, SPP1116, 436POL), 
ESF Programme QUDEDIS, the Russian Foundation for Basic Research,
the Alexander von Humboldt Foundation, and the US Department of Energy.

\begin{appendix}


\section{Calculation of the parameters for the hubbard hamiltonian}
\label{App:Param}

\subsection{Wannier functions}

For periodic boundary conditions the linear part of the Hamiltonian \refer{1HAMIL} 
can be diagonalized in the quasi momentum space using the scheme of Bloch \cite{Ashcroft}. 
In the kagom\'e lattice a single cell contains three equivalent potential
minima, and hence three Wannier functions per unitary cell are required, which can be obtained 
by transforming the Bloch states of the first three bands into the Wannier
basis. 
Let us first consider one particle that is placed in an isolated trimer. 
The Hamiltonian for this model reads
$H = -t \lbrace c^{\dagger}_1 c_2 +c^{\dagger}_2 c_3    +c^{\dagger}_3 c_1+ h.c \rbrace$.
%
%
%
%
The eigenfunctions of $H$ are 
$\vec{a}_0 =  \left( 1, 1, 1 \right)^T /\sqrt{3}$ ,
$\vec{a}_1 = \left(-2, 1, 1 \right)^T/2$ and 
$\vec{a}_2 = \left( 0, -1, 1 \right)^T/\sqrt{2}$ 
with respective eigenvalues $ \lbrace - 2 t  , t , t \rbrace$. 
The transformation matrix $\left( \vec{a}_0 , \vec{a}_1 , \vec{a}_2
\right)^{-1}$  
leads to the states for which just one site of the triangle is occupied.
%
A similar scheme can also be applied for the Bloch functions $\psi^i_k (\r)=\ex{-\im \k \r} u^i_{\k}(\r)$,
where $\k$ is the quasi-momentum, and $u^i_{\k}(\r)$ are the periodic functions of band $i$. 
For a particular quasi momentum $\k$,
the values of $u_{\k}^i(\r)$ at the potential minima are $a_{ij}(\k)$, where
$j\in \{1,2,3\}$ denotes the minimum within the trimer (Fig.~\ref{star}). 
Inverting the matrix $a_{ij}$ for a particular $\k$  
one obtains complex coefficients $c_{op}$, which are then used to
construct a periodic function having its maximum in only one of the three potential minima:
$w_{\k}^o = \sum_{i} c_{oi} u^i_{\k} $, where $o=\{1,2,3\}$ denotes the corners of the trimer.
Similarly as the $u^i_ {\k}$, the $w^o_{\k}$ are functions with the same periodicity as the lattice.
Summation of these functions over $\k$ with a proper phase leads to the Wannier functions \cite{Kohn}:
$ W^o_{\R} =\frac{1}{N_l}       \sum_{\k} \ex{\im \k \R} w^o_{\k}$,
where $\R$ denotes the position of the particular trimer on which
the maximum of the Wannier functions is located.

The point group symmetry of the lattice is broken due to the choice of the
particular set of basis vectors for the reciprocal lattice. A direct consequence of this fact is that 
the Wannier functions within  a trimer cannot be
transformed into each other by a rotation of $\pm 2/3 \pi$ around the center of the trimer.
Hence one obtains different 
hopping probabilities between the sites of the triangle,
$t_{ij}= \langle  W^i_{\R}| H_0 | W^j_{\R} \rangle
        =\frac{1}{N_l^2} \sum_{\k} \sum_{\mu} c_{i,\mu} c_{j,\mu} \epsilon^\mu_{\k} $
where $\epsilon^\mu_{\k}$ is the energy for the quasi momentum $\k$ in band
$\mu$.
%
The Wannier functions can be symmetrized by summing up Bloch functions multiplied
with a $\k$-dependent phase 
factor $\ex{-\im \k \r_i}$, where $\r_i$ is the position of
one of the three potential minima within a cell.
The hopping elements change then to 
$t_{ij}= (1/N_l^2) \sum_{\k} \cos{\k \left(  \r_i - \r_j \right)} \sum_{o} c_{i,o} c_{j,o} \epsilon^o_{\k} $, 
which are independent of the position now. 
The cost of the 
symmetrization is that the Wannier functions are not orthogonal anymore, but 
the overlap remains relatively small.

\subsection{Gaussian  ansatz}

Apart from the case of kagom\'e lattice, it is difficult
to obtain the Wannier functions reliably. 
The coefficients 
for the Hubbard-Hamiltonian can be alternatively obtained using a Gaussian Ansatz \cite{Pedri},
which, in the case of a perfect kagom\'e lattice, and for deep lattice potentials $>5 E_{\rm{rec}}$ 
leads to results which are practically indistinguishable from those of the Wannier functions.
The shape of the Gaussian function reads
$f(x,y) =\sqrt{2/(\sigma_x \sigma_y \pi)} \exp{(-x^2/\sigma_x^2})\exp{(-(y-y_0)^2 /\sigma_y^2)}$.
The center $y_0$, the widths $\sigma_x$ and $\sigma_y$ are the variational parameters minimizing the energy functional:
$
        E=
        \int_{-\infty}^{\infty} \dx 
        \int_{-\infty}^{\infty} \dy 
        \left[\nabla (f(x,y))^2 + f(x,y)^2 V(x,y)\right].
$
The Gauss functions for the other two minima in the trimer 
are obtained by rotating the Gaussian function by $\pm \frac{2}{3} \pi$ around the center of the trimer.

\section{Mean-field theory for a bosonic gas}
\label{App:Meanfield}

The boundaries between Mott-insulator and superfluid phases can be obtained by means 
of a mean-field approach similar as that employed in Ref.~\cite{Lewenstein}. 
We consider only on-site contact-interaction terms, but contrary to Ref.~\cite{Lewenstein} 
we do not restrict ourselves to the hard-core limit.
The system is governed by a Bose-Hubbard Hamiltonian of the form: $H=H_{tr}+H_{hop}$, with 
\begin{eqnarray}
        H_{tr}=
&=&
    - t\sum_{\langle i j \rangle}
          ( b^{\dagger}_i b_j + h. c.)
          \nonumber \\
&&
          + \frac{1}{2} \sum_i n_i (n_i -1) 
           - \mu \sum_i  n_i  
          \quad ,
\label{Htr}
\\
H_{hop}&=&      - t'\sum_{\langle k  l \rangle}
          ( b^{\dagger}_k b_l + h. c.),
\label{Hhop}
\end{eqnarray}
where $t$, $t'$  (denoting the intra- and inter-trimer hoppings) 
and $\mu$ are measured in units of the on-site interaction potential $U$.

Assuming a fixed number of atoms $n$ per trimer, we consider 
all possible Fock-states of the form $|n_1 n_2 n_3 \rangle$ 
with  $n_1+n_2+n_3=n$. For example, for one particle per trimer 
the Hilbert-space contains the Fock-states $|100\rangle$, $|010\rangle$
and $|001\rangle$. Since the model is invariant under 
rotation of $2\pi /3$ the eigenstates are of the form
$|W_1\rangle=(|100\rangle+ z |010\rangle + 
z^2 |001\rangle)/\sqrt{3}$, with $z \in \left\{1,
\exp{(\im \frac{2}{3} \pi)  }, \exp{(-\im \frac{2}{3}} \pi)
\right\}$, implying states with no, left- and 
right-chirality, also known as  W states \cite{Duer}.
We denote 
$z_{\pm}=\exp({\pm \im 2 \pi/3})$ and introduce the operators 
$B_{\pm}= \left( b_1 + z_{\pm} b_2 + z^2_{\pm} b_3 \right)/\sqrt{3}$,
$B_0=\left( b_1 + b_2 + b_3 \right)/\sqrt{3}$.
Their commutation relations are $\left[ B_{\alpha} , B^{\dagger}_{\beta} \right]= \delta_{\alpha \beta}$,
 $\alpha, \beta= \{0,+,-\}$.
The chirality operator is defined as $\chi = (B^{\dagger}_+ B_+ -  B^{\dagger}_- B_-) \mod 3$.
Eq. (\ref{Htr}) can be rewritten into the form 
\begin{eqnarray}
        H_{\rm{tr}}
&=&
        -t 3 B_0^{\dagger} B_0 
        +       
        (t-\mu) \left\{  
                B_0^{\dagger} B_0 + B_+^{\dagger} B_+ + B_-^{\dagger} B_-
        \right\} 
        \nonumber\\
&+&     
        \frac{1}{6} \left\{  
                \left(  {B_0^{\dagger}}^2  + 2 B_+^{\dagger} B_-^{\dagger}  \right)
                \left(  B_0^2 + 2 B_+ B_- \right)
                \right.
                \nonumber\\
&&      
        +       
                \left(  {B_+^{\dagger}}^2  + 2 B_0^{\dagger} B_-^{\dagger}  \right)
                \left(  B_+^2 + 2 B_0 B_- \right)
                \nonumber\\
&&      
                \left.
        +       
                \left(  {B_-^{\dagger}}^2 + 2 B_0^{\dagger} B_+^{\dagger}  \right)
                \left(  B_-^2 + 2 B_0 B_+ \right)
        \right\}
        .
\end{eqnarray}

Therefore $\left[ \exp{ \im 2 \pi /3 (B^{\dagger}_+ B_+-B^{\dagger}_- B_-)},H_{\rm{tr}}\right]=0$,      
and hence the chirality of a state is a conserved quantity. It can be shown 
that the ground state has chirality zero, and therefore we restrict ourselves henceforth to these states. 

For a given number of particles $n$ per trimer, we denote by $|W_n^{\mu}\rangle$ a particular normalized non-chiral 
sum of permutations of a set ${n_1,n_2,  n_3}$, where $\mu$ denotes different allowed non chiral states.
E.g. for two particles per trimer, two possible states are allowed: 
$|W_2^{1}\rangle=(|110\rangle + |101\rangle + |011\rangle )/ \sqrt{3}$
and
$|W_2^{2}\rangle=(|200\rangle + |020\rangle + |002\rangle )/ \sqrt{3}$.
We can then diagonalize the Hamiltonian $H_{tr}$ in this basis, 
$H_n^{\alpha \beta}=\langle W_n^{\alpha}| H_{\rm{tr}} |W_n^{\beta}\rangle$, obtaining the eigenenergies $\epsilon_n^l$ 
and eigenstates $|\psi_n^l\rangle$, where $0\leq l \leq n$.  
The lowest energies $\epsilon^0_n$ for each particle number $n$
have to be compared to obtain the ground-state in the
($t$-$\mu$)-phase space.

If the inter-trimer hopping $t'$ is small, the phase boundaries in the $t$-$t'$-$\mu$ phase diagram can be well estimated 
by using a mean-field approach \cite{sachdev,fisher,stoof2}.We introduce the superfluid order parameter 
$\psi=\langle b_i \rangle=\langle b_i^\dag \rangle$, for every site $i$.
Neglecting fluctuations of $b_i$, $b_j^{\dagger}$ in the second order, we can substitute 
$b_j^\dag b_i=\psi (b_j^\dag +b_i)-\psi ^2$, and hence 
$H_{\rm{hop}}$ can be decomposed into a sum of single-site Hamiltonians of the form:
$
    H_{\rm{hop}}
\approx
    6 t'\psi^2 - 2 \sqrt{3} t' \psi 
    (
        B 
        +
        B^{\dagger}  
    ).
$
The Hamiltonian $H$ can be decomposed then into two parts
$H=H_0 + V$, with $H_0 = H_{\rm{tr}} + 6 t' \psi^2$ and 
$V= -2 t' \sqrt{3} \psi (B_0 + B_0^{\dagger}) $, where
$H_0$ is perturbed by $V$.
Up to second order perturbation theory, the energy becomes of the form 
$E=\epsilon_n^0+r\psi^2$, where 
%
%
%
%
%
%
\begin{equation}
    r= 6 t' \psi^2
        +
         \sum_{m=n\pm1,i}
        \frac{ |\langle \psi_n^0 V  \psi_m^i \rangle |^2 }{\epsilon_n^0- \epsilon_m^i}      .
\label{r}
\end{equation}
The Mott-insulator to superfluid transition may be identified by the equation $r=0$, since 
for $r>0$ the energy is minimized for $\psi^2$ is zero, and for $r<0$ $\psi$ acquires a finite value.
The equation $r=0$ defines a 2D manifold in the $t'$-$t$-$\mu$-parameter space. 

As an example, 
we determine in this Appendix the expression for the boundaries of the Mott-phase with 
one particle per trimer. 
Due to the form of Eq.~(\ref{r}) this calculation demands the knowledge of the eigenenergies and eigenfunctions 
for $n=1$ and $n=2$. For $n=1$, 
$|\psi_1^0\rangle = \left( | 001 \rangle + | 010 \rangle + | 100 \rangle \right)/\sqrt{3}= B^{\dagger} |\psi_0 \rangle$, and 
$\epsilon_1^0=\langle{\psi_1^0} |H_{\rm{tr}}|\psi_1^0 \rangle= -\mu -2 t$. For $n=2$,
\begin{equation} 
\epsilon_2^{0,1}=\frac{1}{2}\left(1 \mp \sqrt{(1 + 2 t)^2  +32 t^2}\right)-t-2\mu,
\end{equation}
and $|\psi_2^{0,1}\rangle=\cos\phi_{0,1} |W_2^2\rangle + \sin\phi_{0,1} |W_2^1\rangle$, 
with 
\begin{equation}
\tan\phi_{0,1}=\frac{1}{4\sqrt{2}t}\{ (1+2t)\mp \sqrt{(1+2t)^2+32t^2}\}.
\end{equation}
At $t'=0$ the region of $1$ particle per trimer is provided by the condition $0\le \epsilon_1^0\le\epsilon_2^0$, i.e. when  
$ -2 t \le \mu \le t +(1 -\sqrt{(1 + 2 t)^2  +32 t^2})/2$. 

After a straightforward but tedious calculation, we can then calculate the sum in Eq.~(\ref{r})
\begin{eqnarray}
&&
        \sum_{m=0,2 ,i}
        \frac{
                |\langle \psi_1^0 |V|\psi_m^i\rangle|^2
        }{
                \epsilon_1^0 - \epsilon_m^i         
        }
        \nonumber \\
&=&
        4 t'^2 \psi^2 \left(  
        \frac{\left(  6\mu - 24 t -4 \right)}{\mu^2 - \mu (2 t +1) - 8 t^2}
        - \frac{3}{2 t + \mu}
        \right)
\end{eqnarray}
Hence, solving for $r=0$, we obtain the value of $t'$ at the phase boundary: 
\begin{eqnarray}
         t' =
        \frac{
        1/2 \left(  \mu^2 - \mu (2t+1) -8 t^2\right) \left(  2 t + \mu\right)
         }{
                (\mu + 8 t) (2 t +1/3) - \mu^2 - 8  t^2
        }.
\end{eqnarray}
\end{appendix}

\end{document}